\documentclass[%
 reprint,
 amsmath,amssymb,
 aps,
]{revtex4-2}
\usepackage{overpic}
\usepackage{makecell}
\usepackage{graphicx}
\usepackage{epsfig}
\usepackage{dcolumn}
\usepackage{bm}
\usepackage{rotating}
\usepackage{multirow}
\usepackage{subfigure}
\usepackage{hhline}
\usepackage{amssymb}

\usepackage{lineno}
\usepackage[colorlinks,linkcolor=blue,anchorcolor=blue,citecolor=blue,urlcolor=blue]{hyperref}
\usepackage{float}
\newcommand\Eone{E_{\rm \uppercase\expandafter{\romannumeral1}}}
\newcommand\Etwo{E_{\rm \uppercase\expandafter{\romannumeral2}}}

\newcommand\bes{BES\uppercase\expandafter{\romannumeral3}}
\newcommand\BEPC{BEPC\uppercase\expandafter{\romannumeral2}}

\let\oldequation\equation
\let\oldendequation\endequation
\renewenvironment{equation}
  {\linenomathNonumbers\oldequation}
  {\oldendequation\endlinenomath}

\let\oldgather\gather
\let\oldendgather\endgather
\renewenvironment{gather}
  {\linenomathNonumbers\oldgather}
  {\oldendgather\endlinenomath}

\begin{document}

\preprint{BESIII/XYZ}

\title{A coupled-channel analysis of the $X(3872)$ lineshape with BESIII data}

\author{
M.~Ablikim$^{1}$, M.~N.~Achasov$^{5,b}$, P.~Adlarson$^{75}$, X.~C.~Ai$^{81}$, R.~Aliberti$^{36}$, A.~Amoroso$^{74A,74C}$, M.~R.~An$^{40}$, Q.~An$^{71,58}$, Y.~Bai$^{57}$, O.~Bakina$^{37}$
% ...
\\
\vspace{0.2cm}
(BESIII Collaboration)\\
\vspace{0.2cm} {\it
$^{1}$ Institute of High Energy Physics, Beijing 100049, People's Republic of China\\
% ...
$^{5}$ Budker Institute of Nuclear Physics SB RAS (BINP), Novosibirsk 630090, Russia\\
% ...
$^{36}$ Johannes Gutenberg University of Mainz, Johann-Joachim-Becher-Weg 45, D-55099 Mainz, Germany\\
$^{37}$ Joint Institute for Nuclear Research, 141980 Dubna, Moscow region, Russia\\
% ...
$^{40}$ Liaoning Normal University, Dalian 116029, People's Republic of China\\
% ...
$^{57}$ Southeast University, Nanjing 211100, People's Republic of China\\
$^{58}$ State Key Laboratory of Particle Detection and Electronics, Beijing 100049, Hefei 230026, People's Republic of China\\
% ...
$^{71}$ University of Science and Technology of China, Hefei 230026, People's Republic of China\\
% ...
$^{74}$ University of Turin and INFN, (A)University of Turin, I-10125, Turin, Italy; (B)University of Eastern Piedmont, I-15121, Alessandria, Italy; (C)INFN, I-10125, Turin, Italy\\
$^{75}$ Uppsala University, Box 516, SE-75120 Uppsala, Sweden\\
% ...
$^{81}$ Zhengzhou University, Zhengzhou 450001, People's Republic of China\\
\vspace{0.2cm}
% ...
$^{b}$ Also at the Novosibirsk State University, Novosibirsk, 630090, Russia\\
% ...
}
}
\noaffiliation{}


\date{\today}

\begin{abstract}

We perform a study of the $X(3872)$ lineshape using the data samples of $e^+e^-\to\gamma X(3872)$, $X(3872)\to D^0\bar{D}^0 \pi^0$ and $\pi^+\pi^- J/\psi$ collected with the \bes~detector. The effects of the coupled-channels and the off-shell $D^{*0}$ are included in the parameterization of the lineshape. The lineshape mass parameter is obtained to be $M_{X}=(3871.63\pm 0.13^{+0.06}_{-0.05})$ MeV. Two poles are found on the first and second Riemann sheets corresponding to the $D^{*0}\bar{D}^0$ branch cut. The pole location on the first sheet is much closer to the $D^{*0}\bar{D}^0$ threshold than the other, and is determined to be $7.04\pm0.15^{+0.07}_{-0.08}$ MeV above the $D^0\bar{D}^0\pi^0$ threshold with an imaginary part $-0.19\pm0.08^{+0.14}_{-0.19}$ MeV. 

\end{abstract}
\maketitle
The $X(3872)$ was discovered in $B^{\pm}\to [X(3872)\to \pi^+\pi^- J/\psi]K^{\pm}$ decay processes by the Belle experiment~\cite{Belle:2003nnu}, and confirmed by CDF~\cite{CDF:2003cab}, D0~\cite{D0:2004zmu} and BaBar~\cite{BaBar:2004oro}. As the first candidate of an exotic charmonium-like state, it has been studied in numerous experimental analyses over the past two decades. 
Along with a well-established quantum number $J^{PC}=1^{++}$~\cite{LHCb:2015jfc}, many remarkable features of the $X(3872)$ have been observed, including a mass almost exactly at the $D^{*0}\bar{D}^0$ threshold, an extremely narrow width~\cite{Workman:2022ynf}, and an isospin-violating decay pattern~\cite{Belle:2005lfc,BaBar:2010wfc,BESIII:2019qvy}. For the nature of the $X(3872)$, many theoretical interpretations have been proposed, including a hadronic molecule~\cite{Swanson:2003tb, Zhao:2014gqa}, a compact tetraquark state~\cite{Maiani:2004vq}, a conventional charmonium state $\chi_{c1}(2P)$~\cite{Achasov:2015oia}, a mixture of a molecule and an excited charmonium state~\cite{Suzuki:2005ha, Kalashnikova:2005ui, Takizawa:2012hy} and so on. However, none of these explanations has been universally accepted.

The lineshape of the $X(3872)$ contains essential information, since from its parameters it is possible to extract the pole locations, the effective range of the particle interaction and the scattering length. Here, the pole of a physical state refers to the corresponding single pole of the off-shell T-matrix in the complex energy plane, where the amplitude becomes infinite.
Recently, the LHCb experiment performed a lineshape study based on a high-statistics $X(3872)\to \pi^+\pi^- J/\psi$ data sample, with both Breit-Wigner and ${\rm Flatt\acute{e}}$ models~\cite{LHCb:2020xds}. However, the lineshapes based on these two models can not be distinguished once the mass resolution is considered. Due to the proximity to the $D^* \bar{D}$ threshold, the lineshape of the $X(3872)$ in the $D^{*0}\bar{D}^0$ channel is significantly distorted, making this channel more sensitive to the behavior of the T-matrix. 
A study of $X(3872)\to D^* \bar{D}$ has previously been performed by Belle~\cite{Belle:2008fma}; however, the off-shell effect of the $D^{*0}$ was not taken into account due to a mass constraint applied to the $D^0\pi^0$ and $D^0\gamma$ systems coming from the $D^{*0}$, which forced the distribution to start from the $D^{*0}\bar{D}^0$ threshold. In the meantime, data samples of both $X(3872)\to\pi^+\pi^-J/\psi$ and $X(3872)\to D^0\bar{D}^0 \pi^0$ channels were acquired by \bes~\cite{ BESIII:2020nbj},
allowing a simultaneous fit, taking into account the coupled-channel effect and the width of the $D^{*0}$, which can improve the $X(3872)$ lineshape measurement.

In this Letter, we present a study of the $X(3872)$ lineshape using $e^+e^-$ annihilation data collected with the BESIII detector at center-of-mass energies ranging from 4.178 to 4.278 GeV, already used in previous  $X(3872)$ studies~\cite{ BESIII:2020nbj}
The total integrated luminosity is 9.0 $\rm fb^{-1}$~\cite{Ablikim2015, Ablikim2022}. The data samples used have center-of-mass energies around the $Y(4230)$ mass peak, since these energies correspond to a maximum in the $X(3872)$ production cross section.
The pole locations of the $X(3872)$ are determined based on a simultaneous fit to the data samples of $X(3872)\to D^0\bar{D}^0 \pi^0$ and $X(3872)\to\pi^+\pi^-  J/\psi$, with the $X(3872)$ produced in the $e^+e^-\to\gamma X(3872)$ process. Throughout this Letter, the charge conjugations are always included and the notations $D^{*0}\bar{D}^0$ and $D^{*+}D^{-}$ denote both themselves and their charge conjugations.

The parameterization scheme in this analysis is developed based on the framework described in Ref.~\cite{Hanhart:2010wh}, taking into account the effects of  the $D^{*0}$ width. In this framework, the $X(3872)$ decays into the three-body final state $D^0\bar{D}^0\pi^0$ via intermediate $D^{*0}\bar{D}^0$. The differential decay rate is written as
\begin{gather}\begin{split}
&\frac{d {\rm Br}(D^0 \bar{D}^0\pi^0)}{d E}={\mathcal B}\frac{{\rm Br}(D^{*0}\to D^0\pi^0)\times g\times k_{\rm eff}(E)}{|D(E)|^2},\\
&\frac{d {\rm Br}(\pi^+\pi^- {J}\slash \psi)}{d E}={\mathcal B}\frac{\Gamma_{\pi^+\pi^-{J}\slash\psi}}{|D(E)|^2},
\end{split}\end{gather}
where the denominator is
\begin{eqnarray}
D(E)=E&-E_X+\frac{1}{2}g\left[\left(\kappa_{\rm eff}(E)+ik^{\phantom{a}}_{\rm eff}(E)\right)\right.\nonumber\\ &+\left.\left(\kappa^c_{\rm eff}(E)+ik^c_{\rm eff}(E)\right)\right]+\frac{i}{2}\Gamma_0. \label{eq:de}
\end{eqnarray}
In the above equations, all the details of the $X(3872)$ production are assumed to be absorbed in a global constant factor ${\mathcal B}$, while ${\rm Br}$ denotes the branching fractions, and $g$ denotes the effective coupling constant of the $X(3872)$ to neutral and charged $D^{*}\bar D$. The energy $E$ ($E_X$) is measured with respect to the three-body $D^0 \bar{D}^0 \pi^0$ threshold, and is related to the invariant mass of the final states (the mass of the $X(3872)$) by $M_{(X)}=m_{D^0}+m_{\bar{D}^0}+m_{\pi^0}+E_{(X)}$, where $m_{D^0}$, $m_{\bar{D}^0}$ and $m_{\pi^0}$ are the masses of $D^0,~\bar{D^0}$ and $\pi^0$ quoted from the Particle Data Group (PDG)~\cite{Workman:2022ynf}. The constant $\Gamma_0$ includes the width of all channels except $D^*\bar{D}$, and is comprised of three parts: $\Gamma_0=\Gamma_{\pi^+\pi^- {J}\slash \psi}+\Gamma_{\rm known}+\Gamma_{\rm unknown} = (1+\beta+\alpha)\Gamma_{\pi^+\pi^- {J}\slash\psi}$. Here, $\Gamma_{\pi^+\pi^- {J}\slash \psi}$, $\Gamma_{\rm known}$ and $\Gamma_{\rm unknown}$ are the partial widths of the $\pi^+\pi^-{J}\slash \psi$ channel, the other measured channels ($\gamma {J}\slash \psi$, $\gamma \psi(3686)$, $\pi^0 \chi_{c1}$, and $\omega {J}\slash\psi$) and the unknown channels, respectively. Due to limited statistics, the ratio $\alpha=\Gamma_{\rm unknown}\slash\Gamma_{\pi^+\pi^- {J}\slash \psi}$ is fixed at $8$ and the ratio $\beta=\Gamma_{\rm known}\slash\Gamma_{\pi^+\pi^- {J}\slash \psi}$ is fixed at $2.8$ according to a global analysis of $X(3872)$ decays~\cite{Li:2019kpj}.
The parameterization of the self-energy terms $\kappa^{(c)}_{\rm eff}(E)$ and $k^{(c)}_{\rm eff}(E)$ (the superscript $c$ indicates the charged $D^{*+}D^-$) can be found in the supplemental material~\cite{sp}.

The expected numbers of signal events in the two decay channels, $\mu_{D^0\bar{D}^0\pi^0}$ and $\mu_{\pi^+\pi^- {J}\slash\psi}$, are related to the number of produced $e^+e^-\to\gamma X(3872)$ events, $\mu_{X(3872)}^{\rm prod}$, as  follows:
\begin{equation}\begin{split}
&\mu_{D^0\bar{D}^0\pi^0}=\epsilon_{D^0\bar{D}^0\pi^0}\times R_{D^0\bar{D}^0\pi^0}\times \mu_{X(3872)}^{\rm prod},\\
&\mu_{\pi^+\pi^- {J}\slash\psi}=\epsilon_{\pi^+\pi^- {J}\slash\psi}\times R_{\pi^+\pi^-{J}\slash\psi}\times\mu^{\rm prod}_{X(3872)}.\label{eq:evtnum}
\end{split}\end{equation}
Here, $\epsilon_{D^0\bar{D}^0\pi^0}$ ($\epsilon_{\pi^+\pi^- {J}\slash\psi}$) represents the efficiency of the $D^0\bar{D}^0\pi^0$ ($\pi^+\pi^- {J}\slash\psi$) channel multiplied by the branching fractions of the decay chains $D^0\to K^-\pi^+,~K^- \pi^+\pi^0,~K^-2\pi^+\pi^-,~\pi^0\to \gamma\gamma$ ($J/\psi\to e^+e^-,~\mu^+\mu^-$), i.e. $1.31\times 10^{-3}$ ($3.78\times 10^{-2}$) according to Refs.~\cite{BESIII:2019qvy, BESIII:2020nbj, Workman:2022ynf}, while $R_{D^0\bar{D}^0\pi^0}$ ($R_{\pi^+\pi^-{J}\slash\psi}$) represents the branching fraction of $X(3872)\to D^0\bar{D}^0\pi^0$ ($X(3872)\to\pi^+\pi^-{J}\slash\psi$) derived from the lineshape analysis. 

The mass resolutions of the two channels are studied based on Monte Carlo (MC) simulation. The MC samples are generated with zero $X(3872)$ width and a series of mass values in the range of interest. For the $D^0\bar{D}^0 \pi^0$ channel, the mass resolution is modeled as a Gaussian function, with a constant mean and a linear mass-dependent width. For the $\pi^+\pi^-J/\psi$ channel, the mass resolution is modeled by a Gaussian function, whose parameters are determined by the control sample $e^+e^-\to \gamma_{\rm ISR}[\psi(2S)\to \pi^+\pi^-J/\psi]$ and calculated at 3.872 GeV.
The values of mass shift and  resolution can be found in the supplemental material~\cite{sp}.
\begin{table}[htbp]
\renewcommand\arraystretch{1.2}
\caption{\label{tab:paras}Summary of the $X(3872)$ lineshape fit parameters.}
\centering \begin{tabular}{ccc}
\hline\hline
Parameter&Symbol&Value\\\hline
Coupling constant&$g$&fit\\
Partial width of $\pi^+\pi^- {J}\slash \psi$&$\Gamma_{\pi^+\pi^- {J}\slash \psi}$&fit\\
Physical mass of $X(3872)$&$M_X$&fit\\
Mass of $D^{*0}$ &-&2.00685 GeV\\
Width of $D^{*0}$&-&55.9 keV\\
Width of $D^{*\pm}$&-&83.4 keV\\
$\Gamma_{{\rm known}}\slash\Gamma_{\pi^+\pi^- {J}\slash \psi}$&$\beta$&2.8\\
$\Gamma_{{\rm unknown}}\slash\Gamma_{\pi^+\pi^- {J}\slash \psi}$&$\alpha$&8\\
\hline
Total number of $X(3872)$ &$\mu^{\rm prod}_{X(3872)}$&fit\\
\multirow{2}*{Efficiency correction\footnote{Multiplied with branching fractions of daughter particles' decay.}} &$\epsilon_{D^0\bar{D}^0\pi^0}$&$1.31\times10^{-3}$~\cite{BESIII:2020nbj}\\
&$\epsilon_{\pi^+\pi^- {J}\slash\psi}$&$3.78\times10^{-2}$~\cite{BESIII:2019qvy}\\
\hline\hline
\end{tabular}

\end{table}

\begin{figure*}[hbtp]
\centering
\subfigure[\label{fig:simufita}]{
\includegraphics[height=4.7cm]{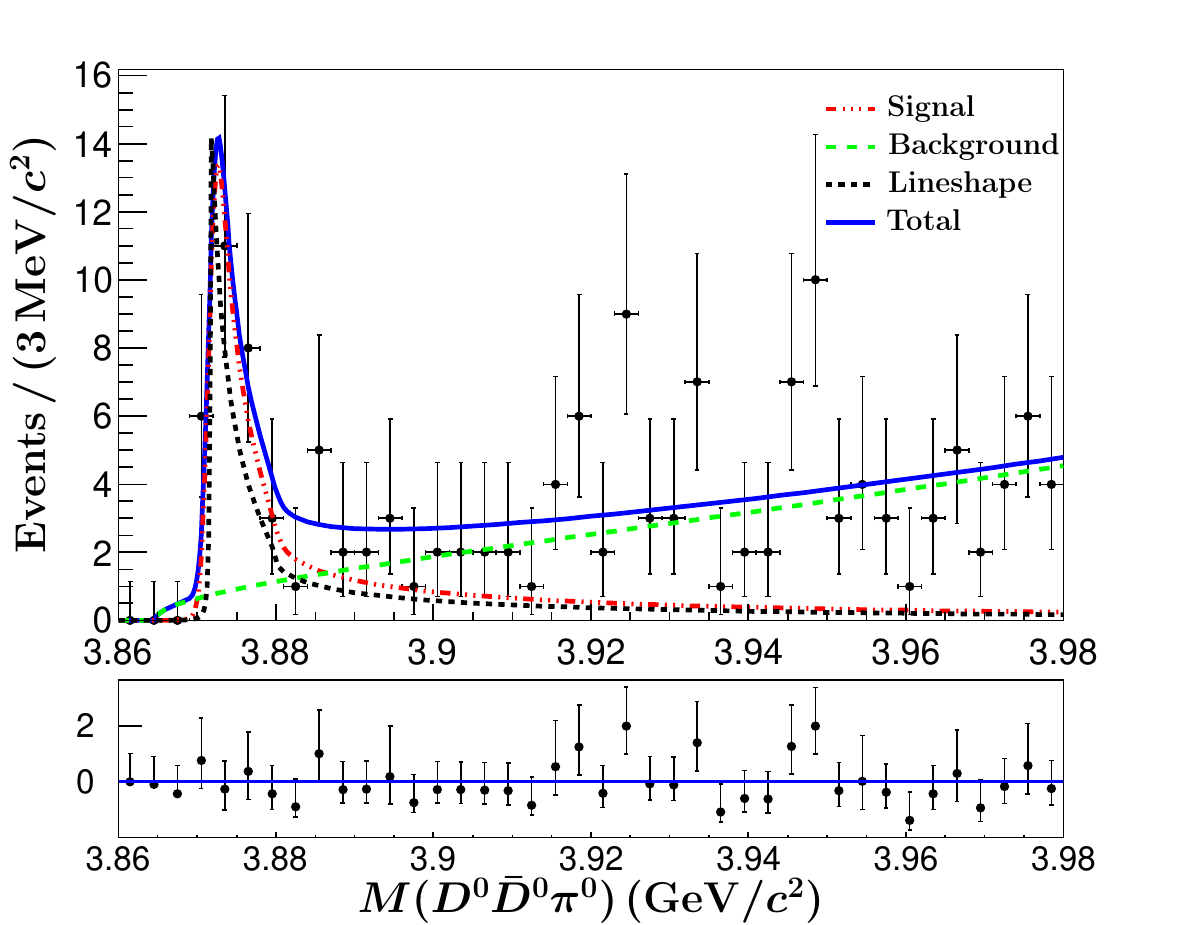}}%
\subfigure[\label{fig:simufitb}]{
\includegraphics[height=4.7cm]{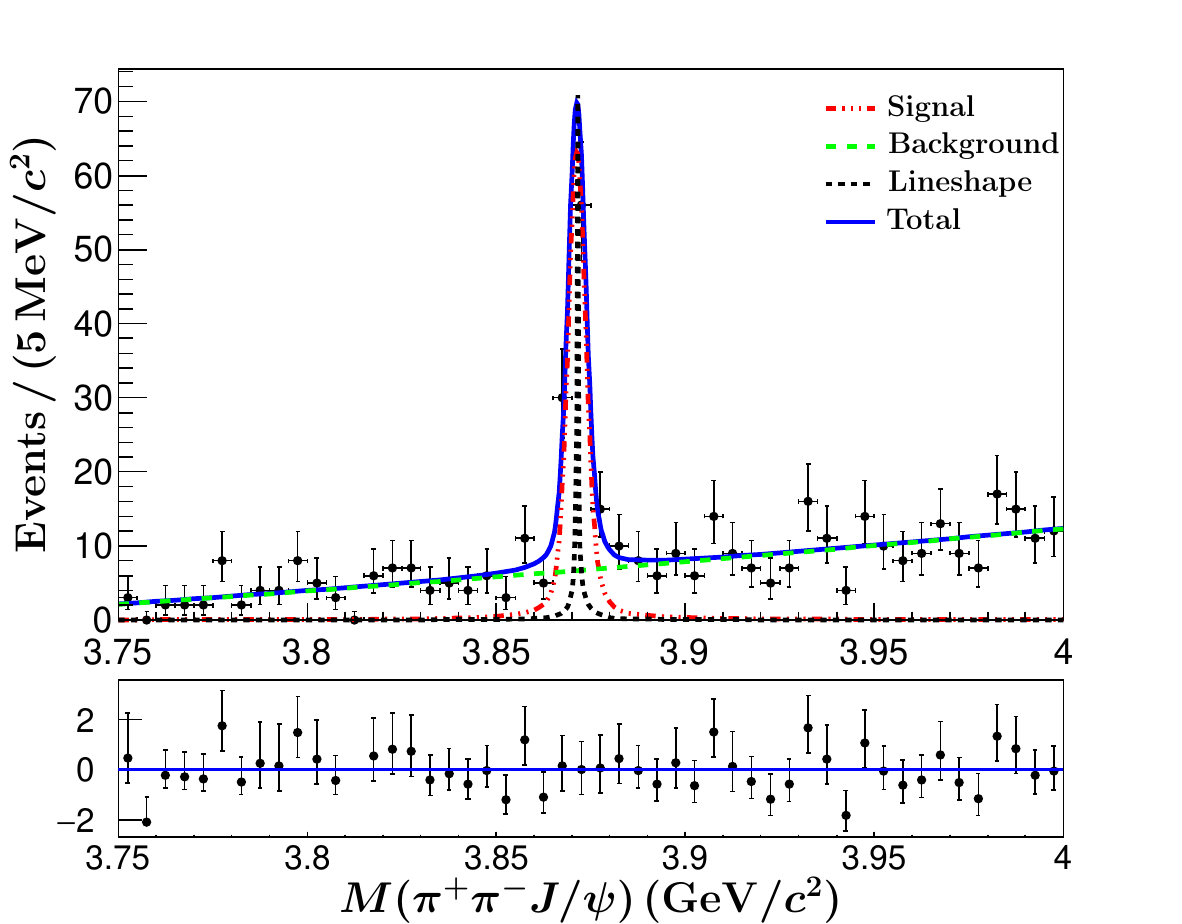}}%
\subfigure[\label{fig:TotalLS}]{\includegraphics[height=4.9cm]{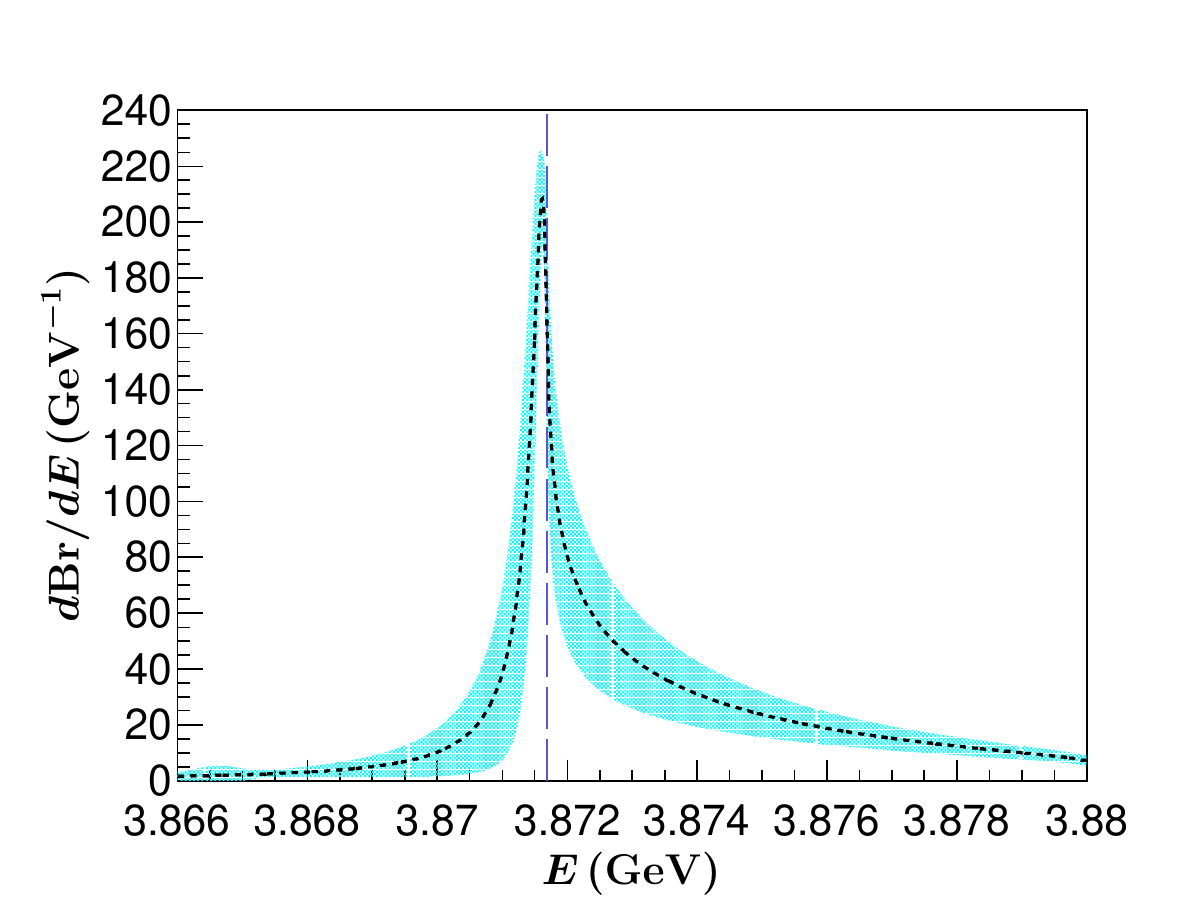}}
\caption{Distributions of (a) $D^0\bar{D}^0\pi^0$  and (b) $\pi^+\pi^- J/\psi$ invariant mass. The black dots with error bars are the data from Ref.~\cite{BESIII:2020nbj}; the blue continuous lines are the probability density functions at the best estimation; the red dotted lines are the signal shapes; the green dashed lines are the background shapes; and the black dashed lines represent the lineshape without the mass resolution considered, normalized to the signal height for comparison. (c) The $X(3872)$ lineshape at the best estimation. The cyan shaded band indicates the statistical uncertainty and the vertical dashed line indicates the position of the $D^{*0} \bar{D}^0 $ threshold.}
\end{figure*}

An unbinned maximum likelihood fit is performed simultaneously to the invariant mass distributions of $M(\pi^+\pi^- J/\psi)$ and $M(D^0\bar{D}^0 \pi^0)$, whose parameters are summarized in Table~\ref{tab:paras}. 
In the fit, to improve the mass resolution, the variable $M(\pi^+\pi^- J/\psi)=M(\pi^+\pi^- l^+l^-)-M(l^+l^-)+m_{J/\psi}$ is used, where $l^+l^-$ stands for $e^+e^-$ or $\mu^+\mu^-$, and $m_{J\slash\psi}$ is the $J\slash\psi$ mass~\cite{Workman:2022ynf}. The background shapes for the $D^0 \bar{D}^0 \pi^0$ and $\pi^+\pi^- J/\psi$ channels are described, respectively, by an ARGUS function~\cite{ALBRECHT1990278} with threshold parameters fixed at the $D^0 \bar{D}^0 \pi^0$ nominal mass, and by a second order Chebyshev function. The obtained lineshape parameters are shown in Table~\ref{tab:corr}, and the number of produced $X(3872)$ is determined to be $\mu_{X(3872)}^{\rm prod}=(9.8\pm3.9)\times 10^4$.
Here, the floating parameter in the fit is $\Gamma_{\pi^+\pi^- J/\psi}$ and its value is transformed into $\Gamma_0=(1+\alpha+\beta)\Gamma_{\pi^+\pi^- J/\psi}$ throughout the letter for convenience. The fit result is shown in Fig.~\ref{fig:simufita} and \ref{fig:simufitb}, and the obtained $X(3872)$ lineshape,  adding all channels together, is shown in Fig.~\ref{fig:TotalLS}, with a full width at the half maximum (FWHM) of 0.44 MeV.

\begin{table}[htbp]
\caption{\label{tab:corr} The fit results of the lineshape parameters and the correlation matrix.}
\renewcommand\arraystretch{1.2}
\centering 
\begin{ruledtabular}
\begin{tabular}{cccc}
Parameters & $g$           & $\Gamma_0$ (MeV) & $M_X$ (MeV)\\\hline
Fit results & $0.16\pm0.10$ & $2.67\pm1.77$    & $3871.63\pm 0.13$\\\hline
$g$        & 1.00          & 0.89             & $-$0.60              \\
$\Gamma_0$ &               & 1.00             & $-$0.29              \\
$M_{X}$    &               &                  & 1.00                 \\
\end{tabular}
\end{ruledtabular}
\end{table}

The systematic uncertainties are estimated as follows.

The uncertainty caused by the choice of the ratios $\alpha = \Gamma_{\rm unknown}\slash\Gamma_{\pi^+\pi^- {J}\slash \psi}$ and $\beta=\Gamma_{\rm known}\slash\Gamma_{\pi^+\pi^- {J}\slash \psi}$ is evaluated by varying $\alpha+\beta$ in the range (4.2,~21.8), according to Ref.~\cite{Li:2019kpj}. 

The $D^{*0}$ nominal width is quoted from an evaluation based on heavy quark symmetry~\cite{Rosner:2013sha}. The corresponding uncertainty is estimated by varying the value in the range (50,~70)~keV in the lineshape models, where the range is determined according to various calculations of the $D^{*0}$ width (55.9~keV in Ref.~\cite{Rosner:2013sha}, 53.7~keV in Ref.~\cite{Du:2021zzh} and 68~keV in Ref.~\cite{Hanhart:2007yq}). 

The relative uncertainty from the efficiency ratios of the $D^0\bar{D}^0\pi^0$ and $\pi^+ \pi^- J/\psi$ decays is assigned to be 10\% according to the uncorrelated uncertainties in Ref.~\cite{BESIII:2019qvy,BESIII:2020nbj}, and is propagated to the lineshape parameters by changing the corresponding values in Eq.~(\ref{eq:evtnum}).

The discrepancy in the $D^0\bar{D}^0\pi^0$ mass resolution between MC simulation and data  (referred to as \textit{Resolution} in Table~\ref{Tab:errpara}) is studied using the control sample $e^+e^-\to [D^{*0}\to D^0\pi^0]\bar{D}^0$. The discrepancy is parameterized as a Gaussian function and extracted from the distribution of $M_{\rm control}=|2p_D+p_{\pi^0}|$, where $p_{\pi^0}$ and $p_D$ denote the four-momentum of the $\pi^0$ and of the $D^0$ decaying from $D^{*0}$, respectively. 
The obtained Gaussian function is convoluted additionally with the lineshape for uncertainty evaluation. For the $\pi^+\pi^- J/\psi$ channel, since the MC invariant mass has been modeled to data by using the control sample, the related systematic uncertainty is treated as negligible. 

For the background models (referred as \textit{Background} in Table~\ref{Tab:errpara}), the uncertainty is evaluated by changing the ARGUS function to a third order polynomial, and changing the order of the Chebyshev function from second to third, respectively. 

The uncertainty of the $D^0$ mass, $50$~keV~\cite{Workman:2022ynf}, is propagated to the lineshape parameters by changing the $D^0$ mass by plus or minus 50~keV in the analysis procedure. 

The center-of-mass energies of the $e^+e^-$ collisions for the datasets used in this work are obtained from a measurement of di-muon events, as described in Ref.~\cite{BESIII:2015zbz}. A common uncertainty of $0.8$~MeV for each dataset is adopted, and it is propagated to the lineshape parameters by changing the values of the center-of-mass energies accordingly when applying the kinematic constraints in the event selection. 

The uncertainties caused by MC simulation configurations (referred as \textit{Simulation} in Table~\ref{Tab:errpara}), including the input cross sections and the generator models used in the decay chains, are evaluated as follows. For the input cross sections of $e^+e^-\to\gamma X(3872)$, the measured cross sections of Ref.~\cite{BESIII:2019qvy} are used, instead of using the $Y(4230)$ lineshape quoted from the PDG; for the $\gamma X(3872)$ angular distribution, it is changed from an E1 transition to a pure S-wave; the $\pi^+\pi^-$ pair in the $\pi^+\pi^- J/\psi$ channel is assumed to come from a $\rho^0$ decay, and for the $\rho J/\psi$ angular distribution from $X(3872)$  the partial wave analysis result of Ref.~\cite{LHCb:2015jfc} is adopted, instead of the original S-wave assumption. 

For each of the above mentioned sources of systematic uncertainties, the largest differences caused by varying the values or modifying the inputs with respect to the nominal values are taken as systematic uncertainties, and are treated as independent. The systematic uncertainties of the lineshape parameters are summarized in Table~\ref{Tab:errpara}, where the last row is obtained by summing each term in quadrature.

\begin{table}[htbp] \renewcommand\arraystretch{1.2}
\caption{\label{Tab:errpara} Systematic uncertainties of the lineshape parameters.}
\centering 
\begin{ruledtabular}
\begin{tabular}{l l l l}
Source        & \multicolumn{1}{c}{$g$}             & \multicolumn{1}{c}{$\Gamma_0$ (MeV)}& \multicolumn{1}{c}{$M_X$ (MeV)}\\\hline
$\alpha$          & $+1.08~-0.10$ & $+6.54~-0.65$ & $+0.05~-0.04$ \\
$\Gamma_{D^{*0}}$ & \multicolumn{1}{c}{$-$} & $+0.05~-0.07$ & \multicolumn{1}{c}{$-$} \\
Efficiency        & $+0.05~-0.03$ & $+0.35~-0.24$ & \multicolumn{1}{c}{$-$} \\
Resolution        & \multicolumn{1}{c}{$-$} & \multicolumn{1}{c}{$\pm0.02$} & \multicolumn{1}{c}{$-$} \\
Background& $+0.05$ & $+0.51~-0.24$ & \multicolumn{1}{c}{$\pm0.01$} \\
$M(D^0)$          & \multicolumn{1}{c}{$-$} & $+0.11~-0.09$ & \multicolumn{1}{c}{$\pm0.03$} \\
$E_{\rm cms}$     & $+0.29$ & $+4.57$ &\multicolumn{1}{r}{$-0.01$} \\
Simulation     & \multicolumn{1}{c}{$\pm0.02$} & \multicolumn{1}{c}{$\pm0.26$} & \multicolumn{1}{c}{$\pm0.01$} \\\hline
Sum               & $+1.12~-0.11$ & $+8.01~-0.82$ & $+0.06~-0.05$ \\
\end{tabular}
\end{ruledtabular}
\end{table}

The analytic structure of the amplitude and the corresponding pole locations are studied by extending the energy $E$ from the real axis to the whole complex plane. According to the simplified form in the supplemental material~\cite{sp}, there are two Riemann sheets with respect to the $D^{*0}\bar{D}^0$ threshold, defined by the sign of the $D^{*0}\bar{D}^0$ self-energy term: 
\begin{equation}\begin{split}
&{\rm sheet~ \uppercase\expandafter{\romannumeral1}}:~ -g\sqrt{-2\left(E-E_R+\frac{i\Gamma_{D^{*0}}}{2}\right)}+i\Gamma_0,\\
&{\rm sheet~ \uppercase\expandafter{\romannumeral2}}:~+g\sqrt{-2\left(E-E_R+\frac{i\Gamma_{D^{*0}}}{2}\right)}+i\Gamma_0.
\end{split}\end{equation}

\begin{figure*}
\centering
\subfigure[sheet \uppercase\expandafter{\romannumeral1}:
$\Eone=7.04-0.19i$~MeV]{
\includegraphics[height=4.2cm]{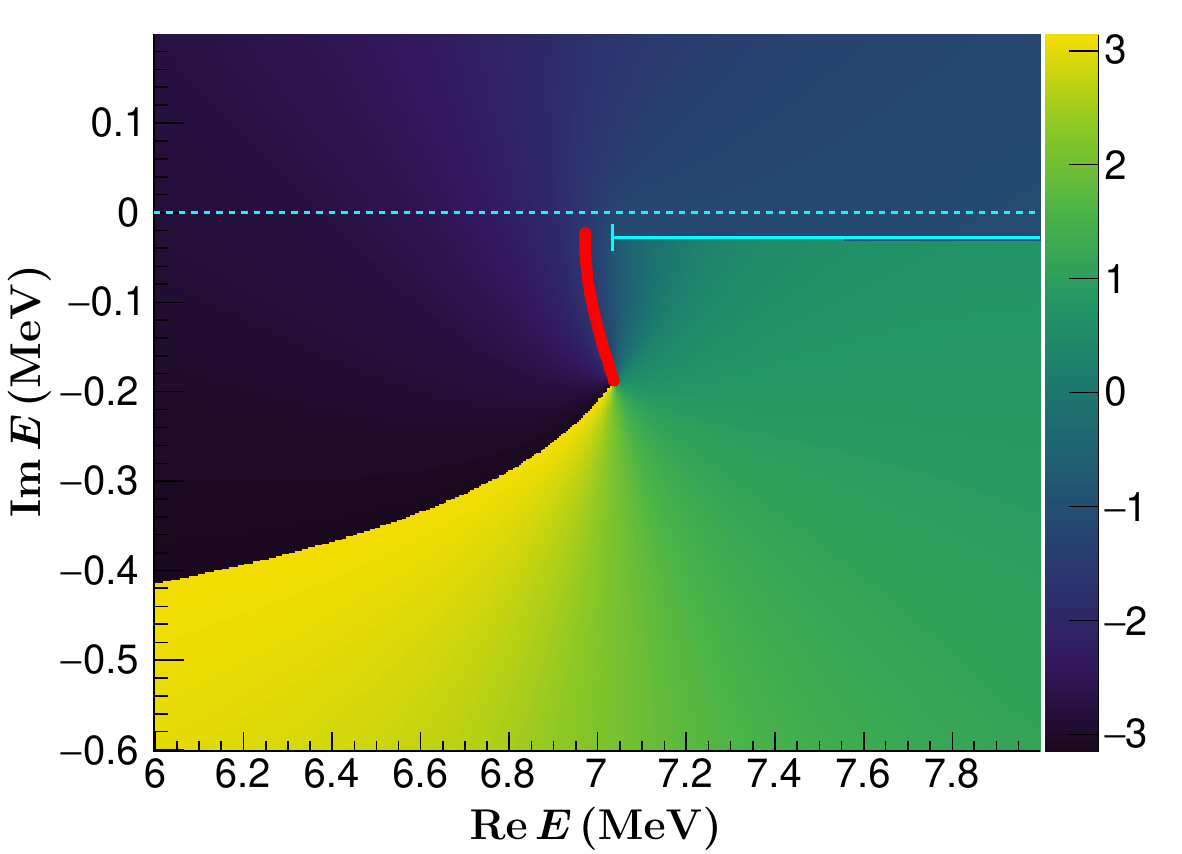}}%
\subfigure[sheet \uppercase\expandafter{\romannumeral2}: $\Etwo=0.26-1.71i$~MeV]{
\includegraphics[height=4.2cm]{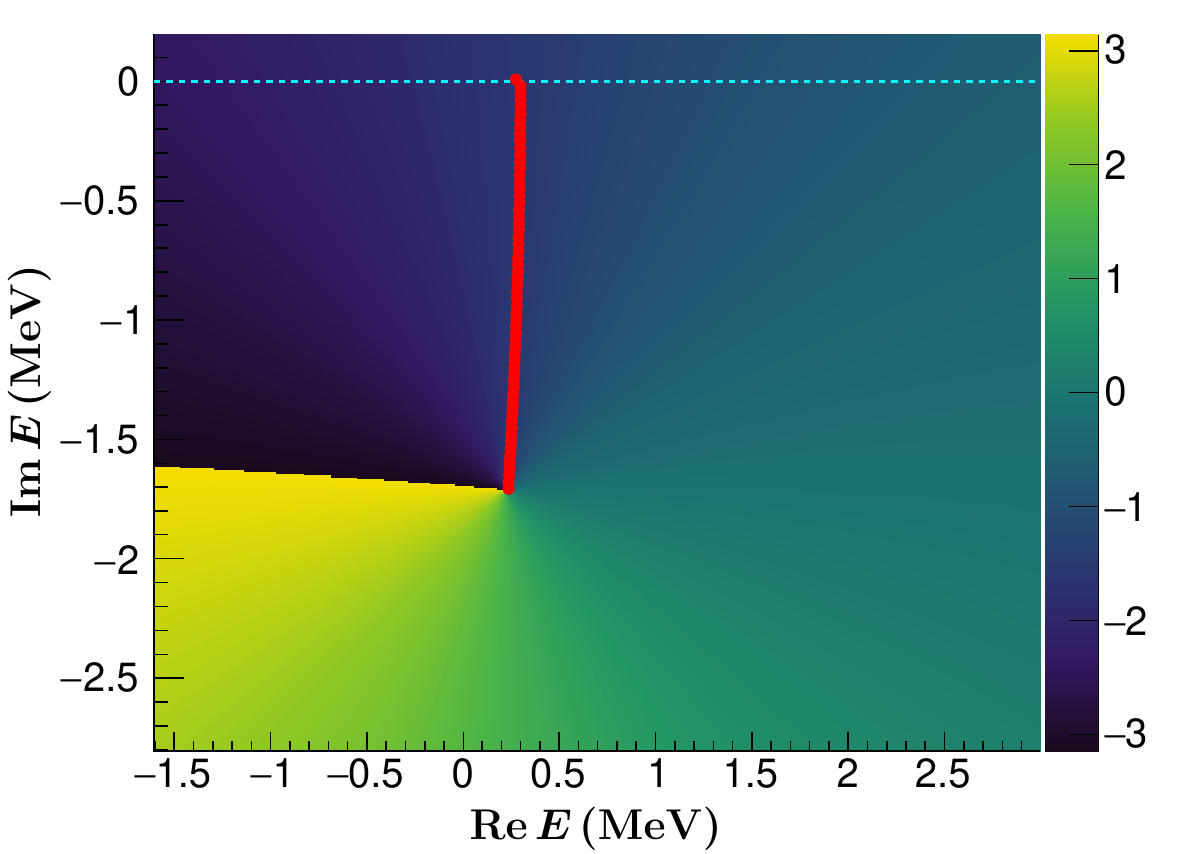}}%
\subfigure[\label{fig:PoleK}$k$-plane: $k^+=-12.6+12.3i$ MeV and $~~~~~~~~~~~k^-=14.1-115.3i$ MeV]{
\includegraphics[height=4.2cm]{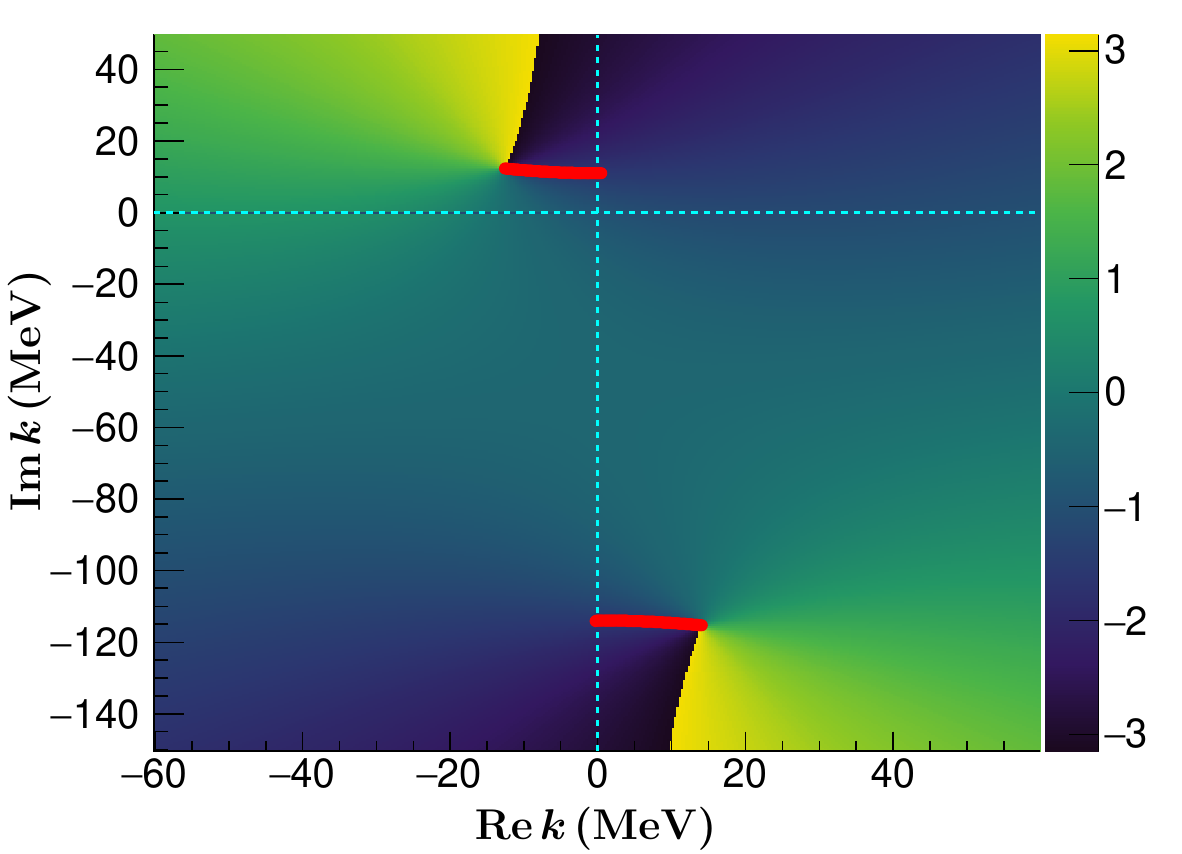}}
\caption{\label{fig:Pole} The phase of the amplitude on (a) {\rm sheet \uppercase\expandafter{\romannumeral1}} and (b) {\rm sheet \uppercase\expandafter{\romannumeral2}} with respect to complex energy. The solid cyan line starting from the point $7.033-0.027i$~MeV is the branch cut, the dashed cyan line is the real axis and the red trajectory approaching the real axis is obtained by continuously decreasing $\Gamma_0$ to 0. (c) The phase of the amplitude with respect to $k$. The upper (lower) half plane corresponds to {\rm sheet \uppercase\expandafter{\romannumeral1}} ({\rm sheet \uppercase\expandafter{\romannumeral2}}).}
\end{figure*}

The numerical results on the pole locations are obtained using a complex roots finding algorithm~\cite{8457320}. The pole locations are visualized by plotting the phase of the amplitude, as shown in Fig.~\ref{fig:Pole}, where the phase is indefinite around the poles and discontinuous across the branch cut. With the nominal lineshape parameters, two poles are found: one is on sheet~\uppercase\expandafter{\romannumeral1}, denoted as $\Eone$, while the other is on sheet~\uppercase\expandafter{\romannumeral2}, denoted as $\Etwo$.  In Fig.~\ref{fig:Pole}, when scaling down $\Gamma_0$ to 0, all channels except $D^*\bar{D}$ switch off, showing that the location of $\Eone$ is much closer to the $D^{*0}\bar{D}^0$ threshold than $\Etwo$.

The pole locations on the $k$-plane are also investigated. The momentum $k$ is defined by $k=\sqrt{2\mu_p}\sqrt{E-E_R+\frac{i\Gamma_{D^{*0}}}{2}}$, where $
\mu_p$ is the two-body reduced mass, and $D(E)$ in Eq.~(\ref{eq:de}) can be rewritten as a function of $k$:
\begin{equation}\label{eq:kplane}
D(k)=\frac{1}{a}-ik+\frac{r_e}{2}k^2+\mathcal{O}(k^3),
\end{equation}
where $a$ is the scattering length and $r_e$ is the effective range~\cite{Hyodo:2013iga}. The high order term occurs due to the presence of charged channels. By doing so, the poles can be displayed in one plane, as shown in Fig.~\ref{fig:PoleK}. The pole location on the upper half plane is $k^+=(-12.6+12.3i)~{\rm MeV}$, and that on the lower half plane is $k^-=(14.1-115.3i)~{\rm MeV}$.

The statistical uncertainties of the pole locations, propagated from the lineshape parameters $g$, $\Gamma_0$, and $M_X$, are obtained by sampling the lineshape parameters according to their covariance matrix. Here, since the uncertainties are large, $g$ and $\Gamma_{\pi^+\pi^- J/\psi}$ could become negative in the $3\sigma$ confidence region, which exceeds the physical boundary. The events with negative $g$ or $\Gamma_{\pi^+\pi^- J/\psi}$ are dropped when calculating the statistical uncertainties of the pole locations (and other related parameters, as described as follows). The systematic uncertainties are obtained using the same treatment as that of the lineshape parameters. The detailed information of statistical and systematic uncertainties can be found in the supplemental material~\cite{sp}. The pole locations including uncertainties are determined to be $\Eone=\left(7.04\pm0.15^{+0.07}_{-0.08}\right)+\left(-0.19\pm0.08^{+0.14}_{-0.19}\right)i$ MeV and $\Etwo=\left(0.26\pm5.74^{+5.14}_{-38.32}\right)+\left(-1.71\pm0.90^{+0.60}_{-1.96}\right)i$ MeV on the Riemann sheets, and $k^+=\left(-12.6\pm5.5^{+6.6}_{-6.2}\right)+\left(12.3\pm6.8^{+6.0}_{-6.4}\right)i$ MeV and $k^-=\left(14.1\pm5.8^{+5.3}_{-2.1}\right)+\left(-115.3\pm44.6^{+52.7}_{-192.8}\right)i$ MeV on the $k$-plane. 

The relative ratio of the branching fractions of $\pi^+\pi^-J/\psi$ and  $D^0\bar{D}^0\pi^0$ is determined to be $\frac{\Gamma(X(3872)\to \pi^+\pi^- J/\psi)}{\Gamma(X(3872)\to D^0 \bar{D}^{*0})}=0.05\pm0.01^{+0.01}_{-0.02}$, which is consistent with the global fit result $0.08\pm0.04$~\cite{Li:2019kpj} and 2020 \bes ~result $0.08\pm0.02$~\cite{BESIII:2020nbj} within $1\sigma$, but lower than the value used as a Gaussian constraint in LHCb's work $0.11\pm0.03$~\cite{LHCb:2020xds}. Compared to the previous result in Ref.~\cite{BESIII:2020nbj}, the ratio is larger mainly due to the inclusion of the $D^{*+}D^-$ term in the model, which 
extends the tail of the lineshape in the $D^0\bar{D}^0\pi^0$ channel and results in a larger signal yield. 

We have estimated the effective range expansion (ERE) parameters, i.e. the scattering length $a$ and the effective range $r_e$ in Eq.~(\ref{eq:kplane}). We consider a simplified case, according to the discussion in Ref.~\cite{Esposito:2021vhu}, by setting $\Gamma_0$ and $\Gamma_{D^{*}}$ to be $0$. The amplitude is reduced to a single channel, $D^*\bar{D}$, and the $D^*$ is treated as a stable particle. After the simplification, the ERE parameters are determined to be $a=\left(-16.5^{+7.0~~+5.6}_{-27.6~-27.7}\right)$~fm  and $r_e=\left(-4.1^{+0.9~+2.8}_{-3.3~-4.4}\right)$~fm. 

Based on the obtained results, we can do a comparison between the $X(3872)$ and the deuteron.
The ERE parameters are related to the field renormalization
constant $Z$ by:

\begin{eqnarray}
a=-\frac{2(1-Z)}{(2-Z)}\frac{1}{\gamma}+\mathcal{O}(\beta^{-1}),\label{eq:Za}\\
r_e=-\frac{Z}{1-Z}\frac{1}{\gamma}+\mathcal{O}(\beta^{-1}).\label{eq:Zre}
\end{eqnarray}
where $\gamma=\sqrt{2\mu_p E_B}$, with $E_B$ the binding energy; the scale $\beta$ measures the momentum scale of the binding interaction, which cannot be calculated without knowing the details of the interaction, but can be estimated to be of the order of the pion mass $m_\pi$ for both neutron-proton and $D^*\bar{D}$
interactions, i.e. $1\slash\beta\sim1\slash m_\pi\simeq1.4~{\rm fm}$.

In the limit of $Z\to0$, the effective range should be positive and dominated by the range correction~\cite{PhysRev.137.B672, BARU2022137290, Matuschek:2020gqe, Esposito:2021vhu}. It was found that this case is compatible with the measured ERE parameters of the deuteron, which is known to be a predominantly molecular state~\cite{PhysRev.137.B672}. However, in the case of the $X(3872)$, we see that a negative effective range can fit the data well, which is different from the deuteron case ($+1.75~{\rm fm}$) and also
suggests an elementary component in the $X(3872)$~\cite{Esposito:2021vhu}. Given that the range correction is less
important in the case of the $X(3872)$, one can solve Eqs.~(\ref{eq:Za}) and (\ref{eq:Zre}) for $Z$ by neglecting $\mathcal{O}(\beta^{-1})$, and obtain $Z=0.18$. However, in the case of the deuteron, it would be impossible to solve Eqs.~(\ref{eq:Za}) and (\ref{eq:Zre}) for $Z$ in a model independent
way since the range correction term is non-negligible for $r_e$. Despite  this, using the
generalized compositeness $\widetilde{X}_A$ proposed in Ref.~\cite{Matuschek:2020gqe}, we find both the $X(3872)$ and the deuteron have similar compositeness. Nevertheless, there are
still large uncertainties in the ERE parameters, which prevent us from drawing strong conclusions on the
nature of the $X(3872)$. More statistics would be helpful. 

In summary, we measure the lineshape of the $X(3872)$ by performing a simultaneous fit to the decay channels $D^0\bar{D}^0\pi^0$ and $\pi^+\pi^- J/\psi$. The lineshape parameters are determined to be $g=0.16\pm0.10^{+1.12}_{-0.11}$, $\Gamma_0=(2.67\pm1.77^{+8.01}_{-0.82})$~MeV and $M_{X}=(3871.63\pm 0.13^{+0.06}_{-0.05})$~MeV, where the first and second uncertainties are statistical and systematical, respectively. The FWHM of the lineshape is determined to be $\left(0.44^{+0.13~+0.38}_{-0.35~-0.25}\right)$~MeV.

The BESIII Collaboration thanks the staff of BEPCII and the IHEP computing center for their strong support. This work is supported in part by National Key R\&D Program of China under Contracts Nos. 2020YFA0406300, 2020YFA0406400; National Natural Science Foundation of China (NSFC) under Contracts Nos. 11635010, 11735014, 11835012, 11935015, 11935016, 11935018, 11961141012, 12022510, 12025502, 12035009, 12035013, 12061131003, 12192260, 12192261, 12192262, 12192263, 12192264, 12192265, 12221005, 12225509, 12235017; the Chinese Academy of Sciences (CAS) Large-Scale Scientific Facility Program; the CAS Center for Excellence in Particle Physics (CCEPP); Joint Large-Scale Scientific Facility Funds of the NSFC and CAS under Contract No. U1832207; CAS Key Research Program of Frontier Sciences under Contracts Nos. QYZDJ-SSW-SLH003, QYZDJ-SSW-SLH040; 100 Talents Program of CAS; The Institute of Nuclear and Particle Physics (INPAC) and Shanghai Key Laboratory for Particle Physics and Cosmology; ERC under Contract No. 758462; European Union's Horizon 2020 research and innovation programme under Marie Sklodowska-Curie grant agreement under Contract No. 894790; German Research Foundation DFG under Contracts Nos. 443159800, 455635585, Collaborative Research Center CRC 1044, FOR5327, GRK 2149; Istituto Nazionale di Fisica Nucleare, Italy; Ministry of Development of Turkey under Contract No. DPT2006K-120470; National Research Foundation of Korea under Contract No. NRF-2022R1A2C1092335; National Science and Technology fund of Mongolia; National Science Research and Innovation Fund (NSRF) via the Program Management Unit for Human Resources \& Institutional Development, Research and Innovation of Thailand under Contract No. B16F640076; Polish National Science Centre under Contract No. 2019/35/O/ST2/02907; The Swedish Research Council; U. S. Department of Energy under Contract No. DE-FG02-05ER41374

\bibliographystyle{bst/apsrev4-2}
\bibliography{REFX3872}

\begin{thebibliography}{35}%
\makeatletter
\providecommand \@ifxundefined [1]{%
 \@ifx{#1\undefined}
}%
\providecommand \@ifnum [1]{%
 \ifnum #1\expandafter \@firstoftwo
 \else \expandafter \@secondoftwo
 \fi
}%
\providecommand \@ifx [1]{%
 \ifx #1\expandafter \@firstoftwo
 \else \expandafter \@secondoftwo
 \fi
}%
\providecommand \natexlab [1]{#1}%
\providecommand \enquote  [1]{``#1''}%
\providecommand \bibnamefont  [1]{#1}%
\providecommand \bibfnamefont [1]{#1}%
\providecommand \citenamefont [1]{#1}%
\providecommand \href@noop [0]{\@secondoftwo}%
\providecommand \href [0]{\begingroup \@sanitize@url \@href}%
\providecommand \@href[1]{\@@startlink{#1}\@@href}%
\providecommand \@@href[1]{\endgroup#1\@@endlink}%
\providecommand \@sanitize@url [0]{\catcode `\\12\catcode `\$12\catcode
  `\&12\catcode `\#12\catcode `\^12\catcode `\_12\catcode `\%12\relax}%
\providecommand \@@startlink[1]{}%
\providecommand \@@endlink[0]{}%
\providecommand \url  [0]{\begingroup\@sanitize@url \@url }%
\providecommand \@url [1]{\endgroup\@href {#1}{\urlprefix }}%
\providecommand \urlprefix  [0]{URL }%
\providecommand \Eprint [0]{\href }%
\providecommand \doibase [0]{https://doi.org/}%
\providecommand \selectlanguage [0]{\@gobble}%
\providecommand \bibinfo  [0]{\@secondoftwo}%
\providecommand \bibfield  [0]{\@secondoftwo}%
\providecommand \translation [1]{[#1]}%
\providecommand \BibitemOpen [0]{}%
\providecommand \bibitemStop [0]{}%
\providecommand \bibitemNoStop [0]{.\EOS\space}%
\providecommand \EOS [0]{\spacefactor3000\relax}%
\providecommand \BibitemShut  [1]{\csname bibitem#1\endcsname}%
\let\auto@bib@innerbib\@empty
\bibitem [{\citenamefont {Choi}\ \emph {et~al.}(2003)\citenamefont {Choi} \emph
  {et~al.}}]{Belle:2003nnu}%
  \BibitemOpen
  \bibfield  {author} {\bibinfo {author} {\bibfnamefont {S.~K.}\ \bibnamefont
  {Choi}} \emph {et~al.} [\bibinfo {collaboration} {Belle Collaboration}],\
  }\href {https://doi.org/10.1103/PhysRevLett.91.262001} {\bibfield  {journal}
  {\bibinfo  {journal} {Phys. Rev. Lett.}\ }\textbf {\bibinfo {volume} {91}},\
  \bibinfo {pages} {262001} (\bibinfo {year} {2003})}\BibitemShut {NoStop}%
\bibitem [{\citenamefont {Acosta}\ \emph {et~al.}(2004)\citenamefont {Acosta}
  \emph {et~al.}}]{CDF:2003cab}%
  \BibitemOpen
  \bibfield  {author} {\bibinfo {author} {\bibfnamefont {D.}~\bibnamefont
  {Acosta}} \emph {et~al.} [\bibinfo {collaboration} {CDF Collaboration}],\
  }\href {https://doi.org/10.1103/PhysRevLett.93.072001} {\bibfield  {journal}
  {\bibinfo  {journal} {Phys. Rev. Lett.}\ }\textbf {\bibinfo {volume} {93}},\
  \bibinfo {pages} {072001} (\bibinfo {year} {2004})}\BibitemShut {NoStop}%
\bibitem [{\citenamefont {Abazov}\ \emph {et~al.}(2004)\citenamefont {Abazov}
  \emph {et~al.}}]{D0:2004zmu}%
  \BibitemOpen
  \bibfield  {author} {\bibinfo {author} {\bibfnamefont {V.~M.}\ \bibnamefont
  {Abazov}} \emph {et~al.} [\bibinfo {collaboration} {D0 Collaboration}],\
  }\href {https://doi.org/10.1103/PhysRevLett.93.162002} {\bibfield  {journal}
  {\bibinfo  {journal} {Phys. Rev. Lett.}\ }\textbf {\bibinfo {volume} {93}},\
  \bibinfo {pages} {162002} (\bibinfo {year} {2004})}\BibitemShut {NoStop}%
\bibitem [{\citenamefont {Aubert}\ \emph {et~al.}(2005)\citenamefont {Aubert}
  \emph {et~al.}}]{BaBar:2004oro}%
  \BibitemOpen
  \bibfield  {author} {\bibinfo {author} {\bibfnamefont {B.}~\bibnamefont
  {Aubert}} \emph {et~al.} [\bibinfo {collaboration} {BaBar Collaboration}],\
  }\href {https://doi.org/10.1103/PhysRevD.71.071103} {\bibfield  {journal}
  {\bibinfo  {journal} {Phys. Rev. D}\ }\textbf {\bibinfo {volume} {71}},\
  \bibinfo {pages} {071103} (\bibinfo {year} {2005})}\BibitemShut {NoStop}%
\bibitem [{\citenamefont {Aaij}\ \emph {et~al.}(2015)\citenamefont {Aaij} \emph
  {et~al.}}]{LHCb:2015jfc}%
  \BibitemOpen
  \bibfield  {author} {\bibinfo {author} {\bibfnamefont {R.}~\bibnamefont
  {Aaij}} \emph {et~al.} [\bibinfo {collaboration} {LHCb Collaboration}],\
  }\href {https://doi.org/10.1103/PhysRevD.92.011102} {\bibfield  {journal}
  {\bibinfo  {journal} {Phys. Rev. D}\ }\textbf {\bibinfo {volume} {92}},\
  \bibinfo {pages} {011102} (\bibinfo {year} {2015})}\BibitemShut {NoStop}%
\bibitem [{\citenamefont {Workman}\ \emph {et~al.}(2022)\citenamefont {Workman}
  \emph {et~al.}}]{Workman:2022ynf}%
  \BibitemOpen
  \bibfield  {author} {\bibinfo {author} {\bibfnamefont {R.~L.}\ \bibnamefont
  {Workman}} \emph {et~al.} [\bibinfo {collaboration} {Particle Data Group}],\
  }\href {https://doi.org/10.1093/ptep/ptac097} {\bibfield  {journal} {\bibinfo
   {journal} {PTEP}\ }\textbf {\bibinfo {volume} {2022}},\ \bibinfo {pages}
  {083C01} (\bibinfo {year} {2022})}\BibitemShut {NoStop}%
\bibitem [{\citenamefont {Abe}\ \emph {et~al.}()\citenamefont {Abe} \emph
  {et~al.}}]{Belle:2005lfc}%
  \BibitemOpen
  \bibfield  {author} {\bibinfo {author} {\bibfnamefont {K.}~\bibnamefont
  {Abe}} \emph {et~al.} [\bibinfo {collaboration} {Belle Collaboration}],\
  }\href@noop {} {\ }\Eprint {https://arxiv.org/abs/hep-ex/0505037}
  {hep-ex/0505037 [hep-ex]} \BibitemShut {NoStop}%
\bibitem [{\citenamefont {del Amo~Sanchez}\ \emph {et~al.}(2010)\citenamefont
  {del Amo~Sanchez} \emph {et~al.}}]{BaBar:2010wfc}%
  \BibitemOpen
  \bibfield  {author} {\bibinfo {author} {\bibfnamefont {P.}~\bibnamefont {del
  Amo~Sanchez}} \emph {et~al.} [\bibinfo {collaboration} {BaBar
  Collaboration}],\ }\href {https://doi.org/10.1103/PhysRevD.82.011101}
  {\bibfield  {journal} {\bibinfo  {journal} {Phys. Rev. D}\ }\textbf {\bibinfo
  {volume} {82}},\ \bibinfo {pages} {011101} (\bibinfo {year}
  {2010})}\BibitemShut {NoStop}%
\bibitem [{\citenamefont {Ablikim}\ \emph {et~al.}(2019)\citenamefont {Ablikim}
  \emph {et~al.}}]{BESIII:2019qvy}%
  \BibitemOpen
  \bibfield  {author} {\bibinfo {author} {\bibfnamefont {M.}~\bibnamefont
  {Ablikim}} \emph {et~al.} [\bibinfo {collaboration} {BESIII Collaboration}],\
  }\href {https://doi.org/10.1103/PhysRevLett.122.232002} {\bibfield  {journal}
  {\bibinfo  {journal} {Phys. Rev. Lett.}\ }\textbf {\bibinfo {volume} {122}},\
  \bibinfo {pages} {232002} (\bibinfo {year} {2019})}\BibitemShut {NoStop}%
\bibitem [{\citenamefont {Swanson}(2004)}]{Swanson:2003tb}%
  \BibitemOpen
  \bibfield  {author} {\bibinfo {author} {\bibfnamefont {E.~S.}\ \bibnamefont
  {Swanson}},\ }\href {https://doi.org/10.1016/j.physletb.2004.03.033}
  {\bibfield  {journal} {\bibinfo  {journal} {Phys. Lett. B}\ }\textbf
  {\bibinfo {volume} {588}},\ \bibinfo {pages} {189} (\bibinfo {year}
  {2004})}\BibitemShut {NoStop}%
\bibitem [{\citenamefont {Zhao}\ \emph {et~al.}(2014)\citenamefont {Zhao},
  \citenamefont {Ma},\ and\ \citenamefont {Zhu}}]{Zhao:2014gqa}%
  \BibitemOpen
  \bibfield  {author} {\bibinfo {author} {\bibfnamefont {L.}~\bibnamefont
  {Zhao}}, \bibinfo {author} {\bibfnamefont {L.}~\bibnamefont {Ma}},\ and\
  \bibinfo {author} {\bibfnamefont {S.-L.}\ \bibnamefont {Zhu}},\ }\href
  {https://doi.org/10.1103/PhysRevD.89.094026} {\bibfield  {journal} {\bibinfo
  {journal} {Phys. Rev. D}\ }\textbf {\bibinfo {volume} {89}},\ \bibinfo
  {pages} {094026} (\bibinfo {year} {2014})}\BibitemShut {NoStop}%
\bibitem [{\citenamefont {Maiani}\ \emph {et~al.}(2005)\citenamefont {Maiani},
  \citenamefont {Piccinini}, \citenamefont {Polosa},\ and\ \citenamefont
  {Riquer}}]{Maiani:2004vq}%
  \BibitemOpen
  \bibfield  {author} {\bibinfo {author} {\bibfnamefont {L.}~\bibnamefont
  {Maiani}}, \bibinfo {author} {\bibfnamefont {F.}~\bibnamefont {Piccinini}},
  \bibinfo {author} {\bibfnamefont {A.~D.}\ \bibnamefont {Polosa}},\ and\
  \bibinfo {author} {\bibfnamefont {V.}~\bibnamefont {Riquer}},\ }\href
  {https://doi.org/10.1103/PhysRevD.71.014028} {\bibfield  {journal} {\bibinfo
  {journal} {Phys. Rev. D}\ }\textbf {\bibinfo {volume} {71}},\ \bibinfo
  {pages} {014028} (\bibinfo {year} {2005})}\BibitemShut {NoStop}%
\bibitem [{\citenamefont {Achasov}\ and\ \citenamefont
  {Rogozina}(2015)}]{Achasov:2015oia}%
  \BibitemOpen
  \bibfield  {author} {\bibinfo {author} {\bibfnamefont {N.~N.}\ \bibnamefont
  {Achasov}}\ and\ \bibinfo {author} {\bibfnamefont {E.~V.}\ \bibnamefont
  {Rogozina}},\ }\href {https://doi.org/10.1142/S0217732315501813} {\bibfield
  {journal} {\bibinfo  {journal} {Mod. Phys. Lett. A}\ }\textbf {\bibinfo
  {volume} {30}},\ \bibinfo {pages} {1550181} (\bibinfo {year}
  {2015})}\BibitemShut {NoStop}%
\bibitem [{\citenamefont {Suzuki}(2005)}]{Suzuki:2005ha}%
  \BibitemOpen
  \bibfield  {author} {\bibinfo {author} {\bibfnamefont {M.}~\bibnamefont
  {Suzuki}},\ }\href {https://doi.org/10.1103/PhysRevD.72.114013} {\bibfield
  {journal} {\bibinfo  {journal} {Phys. Rev. D}\ }\textbf {\bibinfo {volume}
  {72}},\ \bibinfo {pages} {114013} (\bibinfo {year} {2005})}\BibitemShut
  {NoStop}%
\bibitem [{\citenamefont {Kalashnikova}(2005)}]{Kalashnikova:2005ui}%
  \BibitemOpen
  \bibfield  {author} {\bibinfo {author} {\bibfnamefont {Y.~S.}\ \bibnamefont
  {Kalashnikova}},\ }\href {https://doi.org/10.1103/PhysRevD.72.034010}
  {\bibfield  {journal} {\bibinfo  {journal} {Phys. Rev. D}\ }\textbf {\bibinfo
  {volume} {72}},\ \bibinfo {pages} {034010} (\bibinfo {year}
  {2005})}\BibitemShut {NoStop}%
\bibitem [{\citenamefont {Takizawa}\ and\ \citenamefont
  {Takeuchi}(2013)}]{Takizawa:2012hy}%
  \BibitemOpen
  \bibfield  {author} {\bibinfo {author} {\bibfnamefont {M.}~\bibnamefont
  {Takizawa}}\ and\ \bibinfo {author} {\bibfnamefont {S.}~\bibnamefont
  {Takeuchi}},\ }\href {https://doi.org/10.1093/ptep/ptt063} {\bibfield
  {journal} {\bibinfo  {journal} {PTEP}\ }\textbf {\bibinfo {volume} {2013}},\
  \bibinfo {pages} {093D01} (\bibinfo {year} {2013})}\BibitemShut {NoStop}%
\bibitem [{\citenamefont {Aaij}\ \emph {et~al.}(2020)\citenamefont {Aaij} \emph
  {et~al.}}]{LHCb:2020xds}%
  \BibitemOpen
  \bibfield  {author} {\bibinfo {author} {\bibfnamefont {R.}~\bibnamefont
  {Aaij}} \emph {et~al.} [\bibinfo {collaboration} {LHCb Collaboration}],\
  }\href {https://doi.org/10.1103/PhysRevD.102.092005} {\bibfield  {journal}
  {\bibinfo  {journal} {Phys. Rev. D}\ }\textbf {\bibinfo {volume} {102}},\
  \bibinfo {pages} {092005} (\bibinfo {year} {2020})}\BibitemShut {NoStop}%
\bibitem [{\citenamefont {Aushev}\ \emph {et~al.}(2010)\citenamefont {Aushev}
  \emph {et~al.}}]{Belle:2008fma}%
  \BibitemOpen
  \bibfield  {author} {\bibinfo {author} {\bibfnamefont {T.}~\bibnamefont
  {Aushev}} \emph {et~al.} [\bibinfo {collaboration} {Belle Collaboration}],\
  }\href {https://doi.org/10.1103/PhysRevD.81.031103} {\bibfield  {journal}
  {\bibinfo  {journal} {Phys. Rev. D}\ }\textbf {\bibinfo {volume} {81}},\
  \bibinfo {pages} {031103} (\bibinfo {year} {2010})}\BibitemShut {NoStop}%
\bibitem [{\citenamefont {Ablikim}\ \emph {et~al.}(2020)\citenamefont {Ablikim}
  \emph {et~al.}}]{BESIII:2020nbj}%
  \BibitemOpen
  \bibfield  {author} {\bibinfo {author} {\bibfnamefont {M.}~\bibnamefont
  {Ablikim}} \emph {et~al.} [\bibinfo {collaboration} {BESIII Collaboration}],\
  }\href {https://doi.org/10.1103/PhysRevLett.124.242001} {\bibfield  {journal}
  {\bibinfo  {journal} {Phys. Rev. Lett.}\ }\textbf {\bibinfo {volume} {124}},\
  \bibinfo {pages} {242001} (\bibinfo {year} {2020})}\BibitemShut {NoStop}%
\bibitem [{\citenamefont {Ablikim}\ \emph {et~al.}(2015)\citenamefont {Ablikim}
  \emph {et~al.}}]{Ablikim2015}%
  \BibitemOpen
  \bibfield  {author} {\bibinfo {author} {\bibfnamefont {M.}~\bibnamefont
  {Ablikim}} \emph {et~al.} [\bibinfo {collaboration} {BESIII Collaboration}],\
  }\href {https://doi.org/10.1088/1674-1137/39/9/093001} {\bibfield  {journal}
  {\bibinfo  {journal} {Chinese Physics C}\ }\textbf {\bibinfo {volume} {39}},\
  \bibinfo {pages} {093001} (\bibinfo {year} {2015})}\BibitemShut {NoStop}%
\bibitem [{\citenamefont {Ablikim}\ \emph {et~al.}(2022)\citenamefont {Ablikim}
  \emph {et~al.}}]{Ablikim2022}%
  \BibitemOpen
  \bibfield  {author} {\bibinfo {author} {\bibfnamefont {M.}~\bibnamefont
  {Ablikim}} \emph {et~al.} [\bibinfo {collaboration} {BESIII Collaboration}],\
  }\href {https://doi.org/10.1088/1674-1137/ac80b4} {\bibfield  {journal}
  {\bibinfo  {journal} {Chinese Physics C}\ }\textbf {\bibinfo {volume} {46}},\
  \bibinfo {pages} {113002} (\bibinfo {year} {2022})}\BibitemShut {NoStop}%
\bibitem [{\citenamefont {Hanhart}\ \emph {et~al.}(2010)\citenamefont
  {Hanhart}, \citenamefont {Kalashnikova},\ and\ \citenamefont
  {Nefediev}}]{Hanhart:2010wh}%
  \BibitemOpen
  \bibfield  {author} {\bibinfo {author} {\bibfnamefont {C.}~\bibnamefont
  {Hanhart}}, \bibinfo {author} {\bibfnamefont {Y.~S.}\ \bibnamefont
  {Kalashnikova}},\ and\ \bibinfo {author} {\bibfnamefont {A.~V.}\ \bibnamefont
  {Nefediev}},\ }\href {https://doi.org/10.1103/PhysRevD.81.094028} {\bibfield
  {journal} {\bibinfo  {journal} {Phys. Rev. D}\ }\textbf {\bibinfo {volume}
  {81}},\ \bibinfo {pages} {094028} (\bibinfo {year} {2010})}\BibitemShut
  {NoStop}%
\bibitem [{\citenamefont {Li}\ and\ \citenamefont {Yuan}(2019)}]{Li:2019kpj}%
  \BibitemOpen
  \bibfield  {author} {\bibinfo {author} {\bibfnamefont {C.}~\bibnamefont
  {Li}}\ and\ \bibinfo {author} {\bibfnamefont {C.-Z.}\ \bibnamefont {Yuan}},\
  }\href {https://doi.org/10.1103/PhysRevD.100.094003} {\bibfield  {journal}
  {\bibinfo  {journal} {Phys. Rev. D}\ }\textbf {\bibinfo {volume} {100}},\
  \bibinfo {pages} {094003} (\bibinfo {year} {2019})}\BibitemShut {NoStop}%
\bibitem [{sp()}]{sp}%
  \BibitemOpen
  \href@noop {} {\bibinfo {title} {{See Supplemental Material at
  \href{https://docbes3.ihep.ac.cn/DocDB/0011/001171/013/Supplemental\%20material.pdf}{the
  link to be inserted by the editor} for details about lineshape
  parameterization, the determination of the mass resolution and the
  uncertainties of the pole location.}}}\BibitemShut {Stop}%
\bibitem [{\citenamefont {Albrecht}\ \emph {et~al.}(1990)\citenamefont
  {Albrecht} \emph {et~al.}}]{ALBRECHT1990278}%
  \BibitemOpen
  \bibfield  {author} {\bibinfo {author} {\bibfnamefont {H.}~\bibnamefont
  {Albrecht}} \emph {et~al.},\ }\href
  {https://doi.org/https://doi.org/10.1016/0370-2693(90)91293-K} {\bibfield
  {journal} {\bibinfo  {journal} {Physics Letters B}\ }\textbf {\bibinfo
  {volume} {241}},\ \bibinfo {pages} {278} (\bibinfo {year}
  {1990})}\BibitemShut {NoStop}%
\bibitem [{\citenamefont {Rosner}(2013)}]{Rosner:2013sha}%
  \BibitemOpen
  \bibfield  {author} {\bibinfo {author} {\bibfnamefont {J.~L.}\ \bibnamefont
  {Rosner}},\ }\href {https://doi.org/10.1103/PhysRevD.88.034034} {\bibfield
  {journal} {\bibinfo  {journal} {Phys. Rev. D}\ }\textbf {\bibinfo {volume}
  {88}},\ \bibinfo {pages} {034034} (\bibinfo {year} {2013})}\BibitemShut
  {NoStop}%
\bibitem [{\citenamefont {Du}\ \emph {et~al.}(2022)\citenamefont {Du},
  \citenamefont {Baru}, \citenamefont {Dong}, \citenamefont {Filin},
  \citenamefont {Guo}, \citenamefont {Hanhart}, \citenamefont {Nefediev},
  \citenamefont {Nieves},\ and\ \citenamefont {Wang}}]{Du:2021zzh}%
  \BibitemOpen
  \bibfield  {author} {\bibinfo {author} {\bibfnamefont {M.-L.}\ \bibnamefont
  {Du}}, \bibinfo {author} {\bibfnamefont {V.}~\bibnamefont {Baru}}, \bibinfo
  {author} {\bibfnamefont {X.-K.}\ \bibnamefont {Dong}}, \bibinfo {author}
  {\bibfnamefont {A.}~\bibnamefont {Filin}}, \bibinfo {author} {\bibfnamefont
  {F.-K.}\ \bibnamefont {Guo}}, \bibinfo {author} {\bibfnamefont
  {C.}~\bibnamefont {Hanhart}}, \bibinfo {author} {\bibfnamefont
  {A.}~\bibnamefont {Nefediev}}, \bibinfo {author} {\bibfnamefont
  {J.}~\bibnamefont {Nieves}},\ and\ \bibinfo {author} {\bibfnamefont
  {Q.}~\bibnamefont {Wang}},\ }\href
  {https://doi.org/10.1103/PhysRevD.105.014024} {\bibfield  {journal} {\bibinfo
   {journal} {Phys. Rev. D}\ }\textbf {\bibinfo {volume} {105}},\ \bibinfo
  {pages} {014024} (\bibinfo {year} {2022})}\BibitemShut {NoStop}%
\bibitem [{\citenamefont {Hanhart}\ \emph {et~al.}(2007)\citenamefont
  {Hanhart}, \citenamefont {Kalashnikova}, \citenamefont {Kudryavtsev},\ and\
  \citenamefont {Nefediev}}]{Hanhart:2007yq}%
  \BibitemOpen
  \bibfield  {author} {\bibinfo {author} {\bibfnamefont {C.}~\bibnamefont
  {Hanhart}}, \bibinfo {author} {\bibfnamefont {Y.~S.}\ \bibnamefont
  {Kalashnikova}}, \bibinfo {author} {\bibfnamefont {A.~E.}\ \bibnamefont
  {Kudryavtsev}},\ and\ \bibinfo {author} {\bibfnamefont {A.~V.}\ \bibnamefont
  {Nefediev}},\ }\href {https://doi.org/10.1103/PhysRevD.76.034007} {\bibfield
  {journal} {\bibinfo  {journal} {Phys. Rev. D}\ }\textbf {\bibinfo {volume}
  {76}},\ \bibinfo {pages} {034007} (\bibinfo {year} {2007})}\BibitemShut
  {NoStop}%
\bibitem [{\citenamefont {Ablikim}\ \emph {et~al.}(2016)\citenamefont {Ablikim}
  \emph {et~al.}}]{BESIII:2015zbz}%
  \BibitemOpen
  \bibfield  {author} {\bibinfo {author} {\bibfnamefont {M.}~\bibnamefont
  {Ablikim}} \emph {et~al.} [\bibinfo {collaboration} {BESIII Collaboration}],\
  }\href {https://doi.org/10.1088/1674-1137/40/6/063001} {\bibfield  {journal}
  {\bibinfo  {journal} {Chin. Phys. C}\ }\textbf {\bibinfo {volume} {40}},\
  \bibinfo {pages} {063001} (\bibinfo {year} {2016})}\BibitemShut {NoStop}%
\bibitem [{\citenamefont {Kowalczyk}(2018)}]{8457320}%
  \BibitemOpen
  \bibfield  {author} {\bibinfo {author} {\bibfnamefont {P.}~\bibnamefont
  {Kowalczyk}},\ }\href {https://doi.org/10.1109/TAP.2018.2869213} {\bibfield
  {journal} {\bibinfo  {journal} {IEEE Transactions on Antennas and
  Propagation}\ }\textbf {\bibinfo {volume} {66}},\ \bibinfo {pages} {7198}
  (\bibinfo {year} {2018})}\BibitemShut {NoStop}%
\bibitem [{\citenamefont {Hyodo}(2013)}]{Hyodo:2013iga}%
  \BibitemOpen
  \bibfield  {author} {\bibinfo {author} {\bibfnamefont {T.}~\bibnamefont
  {Hyodo}},\ }\href {https://doi.org/10.1103/PhysRevLett.111.132002} {\bibfield
   {journal} {\bibinfo  {journal} {Phys. Rev. Lett.}\ }\textbf {\bibinfo
  {volume} {111}},\ \bibinfo {pages} {132002} (\bibinfo {year}
  {2013})}\BibitemShut {NoStop}%
\bibitem [{\citenamefont {Esposito}\ \emph {et~al.}(2022)\citenamefont
  {Esposito}, \citenamefont {Maiani}, \citenamefont {Pilloni}, \citenamefont
  {Polosa},\ and\ \citenamefont {Riquer}}]{Esposito:2021vhu}%
  \BibitemOpen
  \bibfield  {author} {\bibinfo {author} {\bibfnamefont {A.}~\bibnamefont
  {Esposito}}, \bibinfo {author} {\bibfnamefont {L.}~\bibnamefont {Maiani}},
  \bibinfo {author} {\bibfnamefont {A.}~\bibnamefont {Pilloni}}, \bibinfo
  {author} {\bibfnamefont {A.~D.}\ \bibnamefont {Polosa}},\ and\ \bibinfo
  {author} {\bibfnamefont {V.}~\bibnamefont {Riquer}},\ }\href
  {https://doi.org/10.1103/PhysRevD.105.L031503} {\bibfield  {journal}
  {\bibinfo  {journal} {Phys. Rev. D}\ }\textbf {\bibinfo {volume} {105}},\
  \bibinfo {pages} {L031503} (\bibinfo {year} {2022})}\BibitemShut {NoStop}%
\bibitem [{\citenamefont {Weinberg}(1965)}]{PhysRev.137.B672}%
  \BibitemOpen
  \bibfield  {author} {\bibinfo {author} {\bibfnamefont {S.}~\bibnamefont
  {Weinberg}},\ }\href {https://doi.org/10.1103/PhysRev.137.B672} {\bibfield
  {journal} {\bibinfo  {journal} {Phys. Rev.}\ }\textbf {\bibinfo {volume}
  {137}},\ \bibinfo {pages} {B672} (\bibinfo {year} {1965})}\BibitemShut
  {NoStop}%
\bibitem [{\citenamefont {Baru}\ \emph {et~al.}(2022)\citenamefont {Baru},
  \citenamefont {Dong}, \citenamefont {Du}, \citenamefont {Filin},
  \citenamefont {Guo}, \citenamefont {Hanhart}, \citenamefont {Nefediev},
  \citenamefont {Nieves},\ and\ \citenamefont {Wang}}]{BARU2022137290}%
  \BibitemOpen
  \bibfield  {author} {\bibinfo {author} {\bibfnamefont {V.}~\bibnamefont
  {Baru}}, \bibinfo {author} {\bibfnamefont {X.-K.}\ \bibnamefont {Dong}},
  \bibinfo {author} {\bibfnamefont {M.-L.}\ \bibnamefont {Du}}, \bibinfo
  {author} {\bibfnamefont {A.}~\bibnamefont {Filin}}, \bibinfo {author}
  {\bibfnamefont {F.-K.}\ \bibnamefont {Guo}}, \bibinfo {author} {\bibfnamefont
  {C.}~\bibnamefont {Hanhart}}, \bibinfo {author} {\bibfnamefont
  {A.}~\bibnamefont {Nefediev}}, \bibinfo {author} {\bibfnamefont
  {J.}~\bibnamefont {Nieves}},\ and\ \bibinfo {author} {\bibfnamefont
  {Q.}~\bibnamefont {Wang}},\ }\href
  {https://doi.org/https://doi.org/10.1016/j.physletb.2022.137290} {\bibfield
  {journal} {\bibinfo  {journal} {Physics Letters B}\ }\textbf {\bibinfo
  {volume} {833}},\ \bibinfo {pages} {137290} (\bibinfo {year}
  {2022})}\BibitemShut {NoStop}%
\bibitem [{\citenamefont {Matuschek}\ \emph {et~al.}(2021)\citenamefont
  {Matuschek}, \citenamefont {Baru}, \citenamefont {Guo},\ and\ \citenamefont
  {Hanhart}}]{Matuschek:2020gqe}%
  \BibitemOpen
  \bibfield  {author} {\bibinfo {author} {\bibfnamefont {I.}~\bibnamefont
  {Matuschek}}, \bibinfo {author} {\bibfnamefont {V.}~\bibnamefont {Baru}},
  \bibinfo {author} {\bibfnamefont {F.-K.}\ \bibnamefont {Guo}},\ and\ \bibinfo
  {author} {\bibfnamefont {C.}~\bibnamefont {Hanhart}},\ }\href
  {https://doi.org/10.1140/epja/s10050-021-00413-y} {\bibfield  {journal}
  {\bibinfo  {journal} {Eur. Phys. J. A}\ }\textbf {\bibinfo {volume} {57}},\
  \bibinfo {pages} {101} (\bibinfo {year} {2021})},\ \Eprint
  {https://arxiv.org/abs/2007.05329} {arXiv:2007.05329 [hep-ph]} \BibitemShut
  {NoStop}%
\end{thebibliography}%

\end{document}


\textbf{\large\boldmath\centerline {Supplemental material}}
\section{Self-energy terms parameterization}
The parameterizations of the self-energy terms $\kappa_{\rm eff}(E)$ and $k_{\rm eff}(E)$ corresponding to the coupling with the $D^{*0}\bar{D}^0$ channel are the following: 
\begin{gather}\begin{split}
k_{\rm eff}(E)=&\sqrt{\mu_p}\sqrt{\sqrt{(E-E_R)^2+\Gamma_{D^{*0}}^2\slash 4}+E-E_R},\\
\kappa_{\rm eff}(E)=&-\sqrt{\mu_p}\sqrt{\sqrt{(E-E_R)^2+\Gamma_{D^{*0}}^2\slash 4}-E+E_R}\\
&+\sqrt{\mu_p}\sqrt{\sqrt{(E_X-E_R)^2+\Gamma_{D^{*0},~X}^2\slash 4}-E_X+E_R}.\label{eq:sfenergy}
\end{split}\end{gather}
The energy-dependent width $\Gamma_{D^{*0}}=\Gamma_{D^{*0}}(E)$ is expressed as:
\begin{equation}\begin{split}
\Gamma_{D^{*0}}(E)=\Gamma_{D^{*0}}^{0}\times\left({\rm Br}(D^{*0}\to D^0\pi^0)\times\left(\frac{E}{E_R}\right)^{3\slash2}+{\rm Br}(D^{*0}\to \gamma D^0)\right),
\end{split}\end{equation}
and $\Gamma_{D^{*0},~X}=\Gamma_{D^{*0}}(E_X)$, where $\Gamma_{D^{*0}}^{0}=55.9$ keV is the nominal width of $D^{*0}$, $E_R=m_{D^{*0}}-m_{D^0}-m_{\pi^0}$, and the two-body reduced mass is $
\mu_p=\frac{m_{D^{*0}}m_{\bar{D}^0}}{m_{D^{*0}}+m_{\bar{D}^0}}.
$

The terms $\kappa^c_{\rm eff}(E)$ and $k^c_{\rm eff}(E)$ corresponding to the charged $D^{*+}D^-$ are obtained by the following replacements in equations~(\ref{eq:sfenergy}): $\mu_p\to \mu_p^c$, $\Gamma_{D^{*0},~(X)}(E)\to\Gamma_{D^{*+},~(X)}(E)$, $E_R\to E_R^c$, where the width of the $D^{*+}$ is
\begin{equation}
\Gamma_{D^{*+}}(E)=\Gamma_{D^{*+}}^{0}\times\left({\rm Br}(D^{*+}\to D^+\pi^0)\times\left(\frac{E_{D^+D^-\pi^0}}{E_R^c}\right)^{\frac32} + {\rm Br}(D^{*+}\to D^0\pi^+)\times\left(\frac{E_{D^0D^-\pi^+}}{E_R^c}\right)^{\frac32}+{\rm Br}(D^{*+}\to D^+\gamma)\right),
\end{equation}
and $\Gamma_{D^{*+},~X}=\Gamma_{D^{*+}}(E_X)$, where $\Gamma_{D^{*+}}^{0}=83.4$ keV is the nominal width of the $D^{*+}$, $E_R^c=m_{D^{*+}}+m_{D^-}-(m_{D^0}+m_{\bar{D}^0}+m_{\pi^0})$, $\mu_p^c=(m_{D^{*+}}m_{D^-})\slash(m_{D^{*+}}+m_{D^-})$, $E_{D^+D^-\pi^0}=M-m_{D^+}-m_{D^-}-m_{\pi^0}$, and $E_{D^0D^-\pi^+}=M-m_{D^0}-m_{D^-}-m_{\pi^+}$. 

For convenience, for the pole search the self-energy terms can be simplified using the following identity:
\begin{equation}
\sqrt{\sqrt{\left(E-E_R\right)^2+\frac{\Gamma_{D^{*0}}^2}{4}}-E+E_R}-i\sqrt{\sqrt{\left(E-E_R\right)^2+\frac{\Gamma_{D^{*0}}^2}{4}}+E-E_R}=\sqrt{-2\left(E-E_R+\frac{i\Gamma_{D^{*0}}}{2}\right)},
\end{equation}
and the self-energy terms become
\begin{equation}
\kappa_{eff}+ik_{eff}=\sqrt{\mu_p}*\left(-\sqrt{-2\left(E-E_R+\frac{i\Gamma_{D^{*0}}}{2}\right)}+\sqrt{\sqrt{\left(E_X-E_R\right)^2+\frac{\Gamma_{D^{*0},~X}^2}{4}}-E_X+E_R}\right).
\end{equation}

\section{DETERMINATION OF THE MASS RESOLUTION}
\begin{figure}[htbp]\centering
\subfigure[]{
\includegraphics[height=6cm]{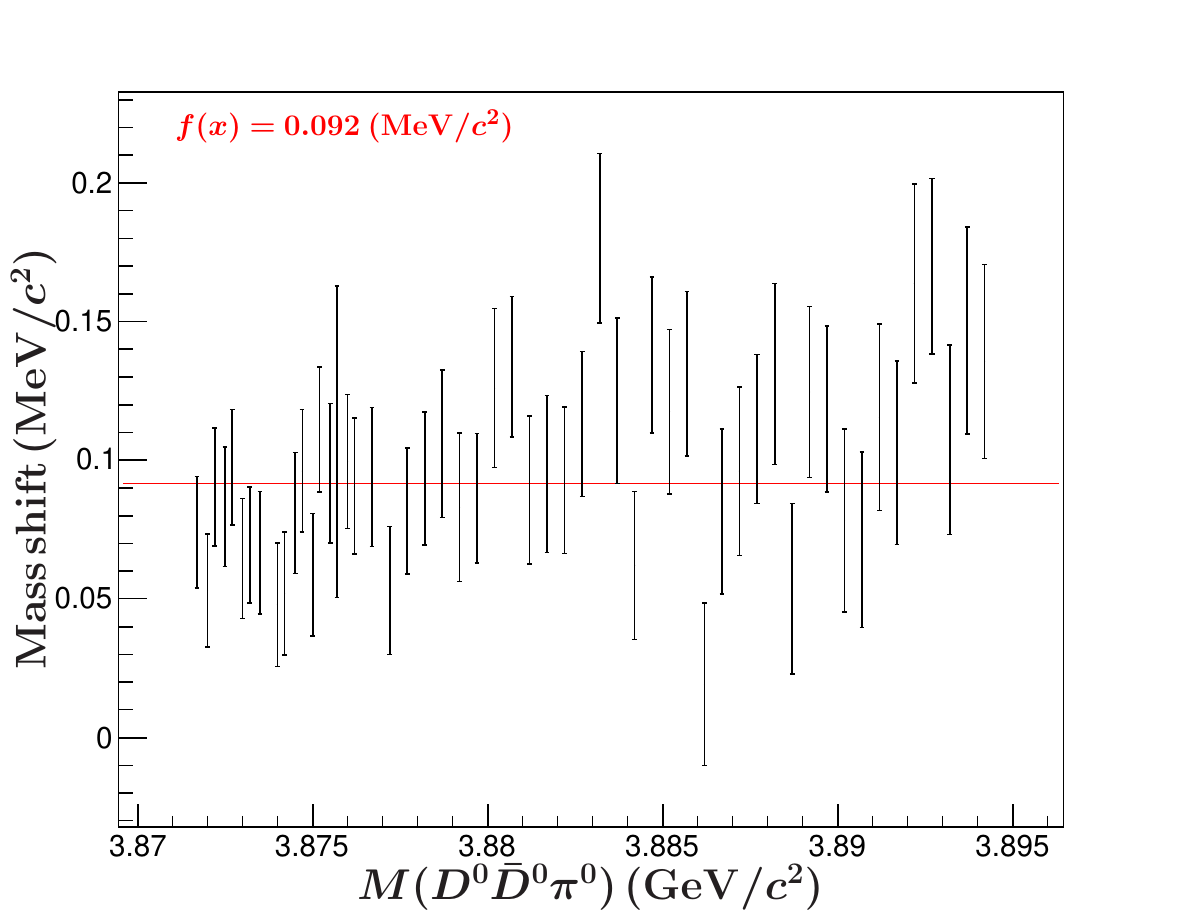}}
\subfigure[]{
\includegraphics[height=6cm]{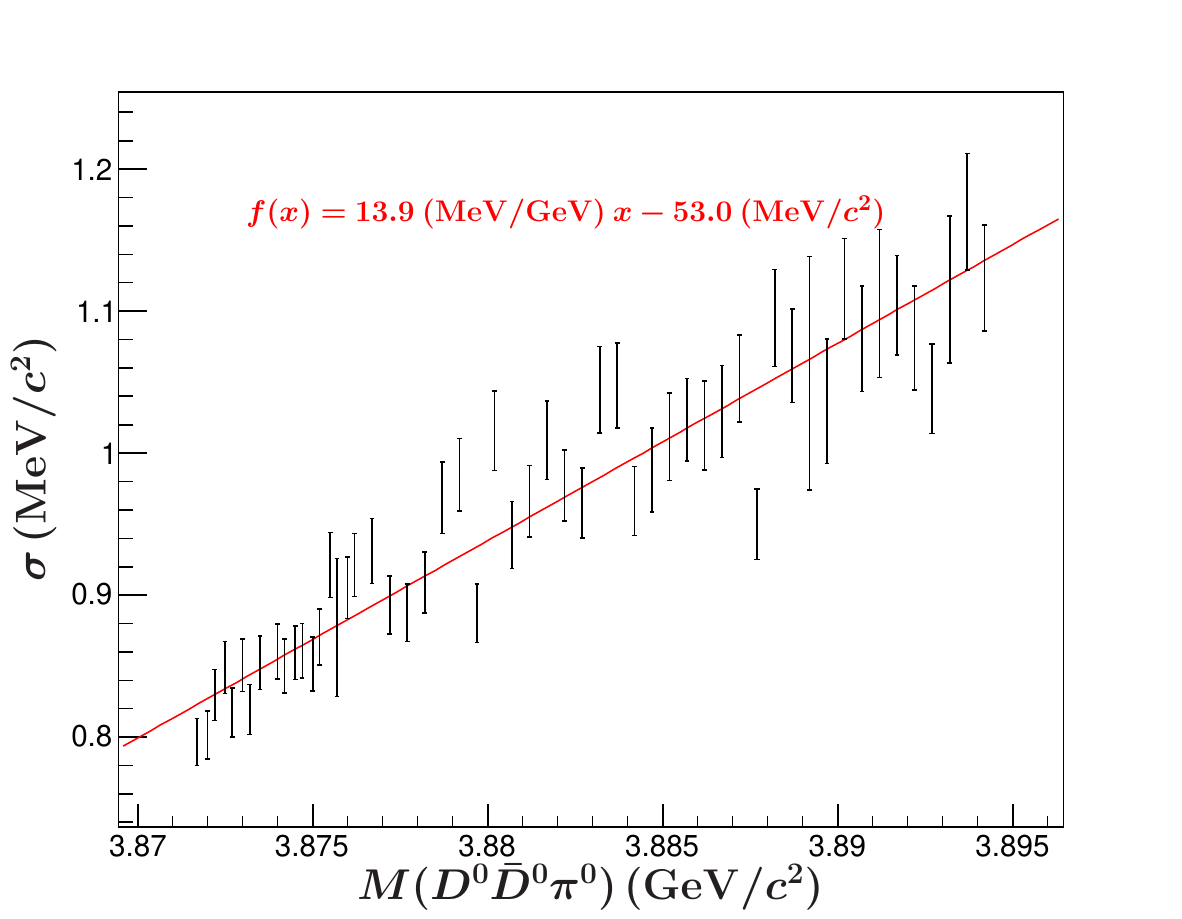}}
\caption{\label{fig:RES}The (a) mass shift and (b) mass resolution of $D^0\bar{D}^0\pi^0$. The red lines are the fit results.}
\end{figure}

\begin{figure}[htbp]\centering
\subfigure[]{
\includegraphics[height=6cm]{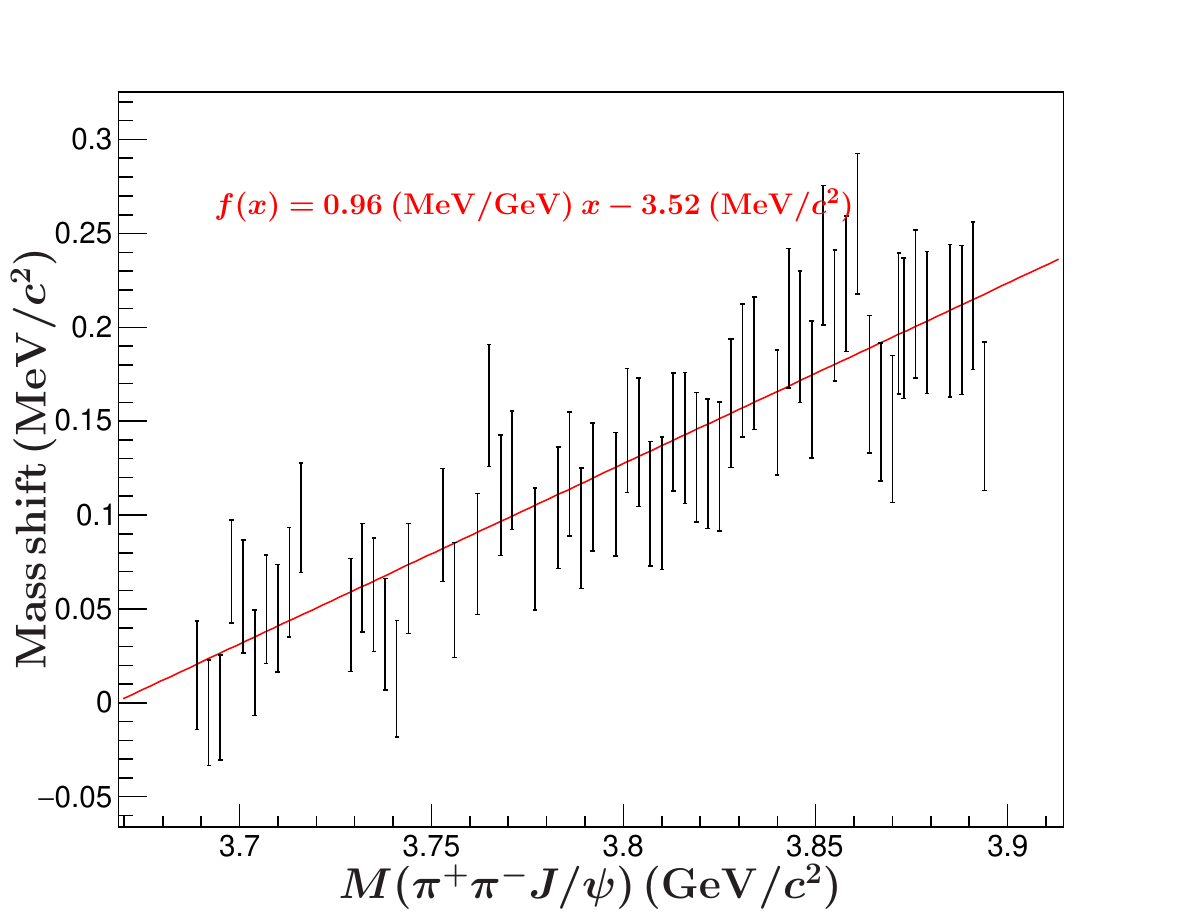}}
\subfigure[]{
\includegraphics[height=6cm]{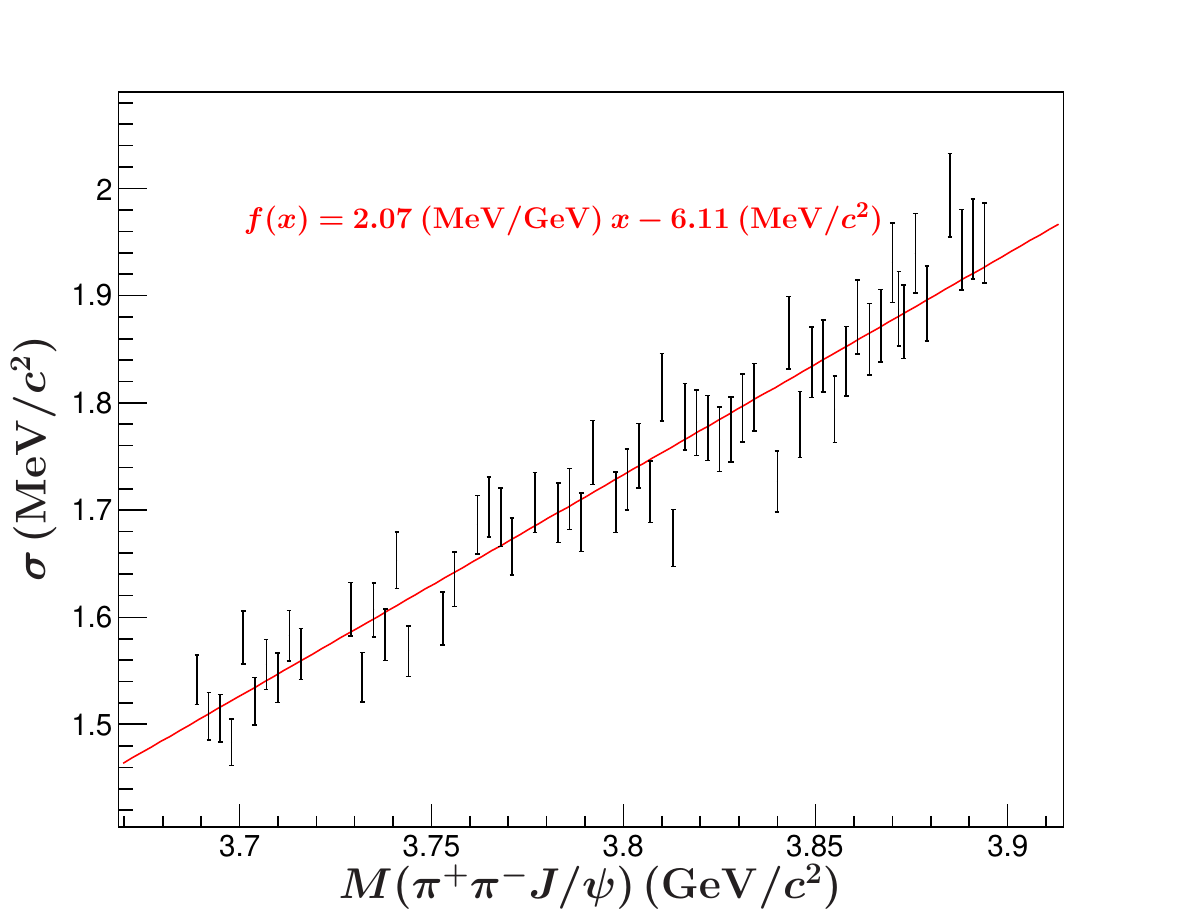}}
\caption{\label{fig:RESpipi} The MC simulation result of the energy dependence of (a) the mass shift and (b) the resolution. The red lines are the fit results.}
\end{figure}

To determine the mass resolution, MC samples are generated. For the $D^0\bar{D}^0\pi^0$ channel, samples of $e^+e^-\to \gamma [X(3872)\to D^0\bar{D}^0\pi^0]$ are generated with zero $X(3872)$ width and values of mass ranging from the threshold to 30 MeV above. For the $\pi^+\pi^-J\slash\psi$ channel, samples of $e^+e^-\to \gamma [X(3872)\to \pi^+\pi^-J\slash\psi]$ are generated with zero $X(3872)$ width and values of mass ranging from 3.686 GeV to 3.872 GeV. The MC simulation results are shown in Fig.~\ref{fig:RES} and \ref{fig:RESpipi}. 

\section{\boldmath SUMMARY OF THE STATISTICAL AND SYSTEMATIC UNCERTAINTIES ON THE POLE LOCATIONS}
\begin{figure}[htbp]
\centering
\subfigure[~sheet \uppercase\expandafter{\romannumeral1}: $7.04-0.19i$ MeV]{
\includegraphics[height=4.6cm]{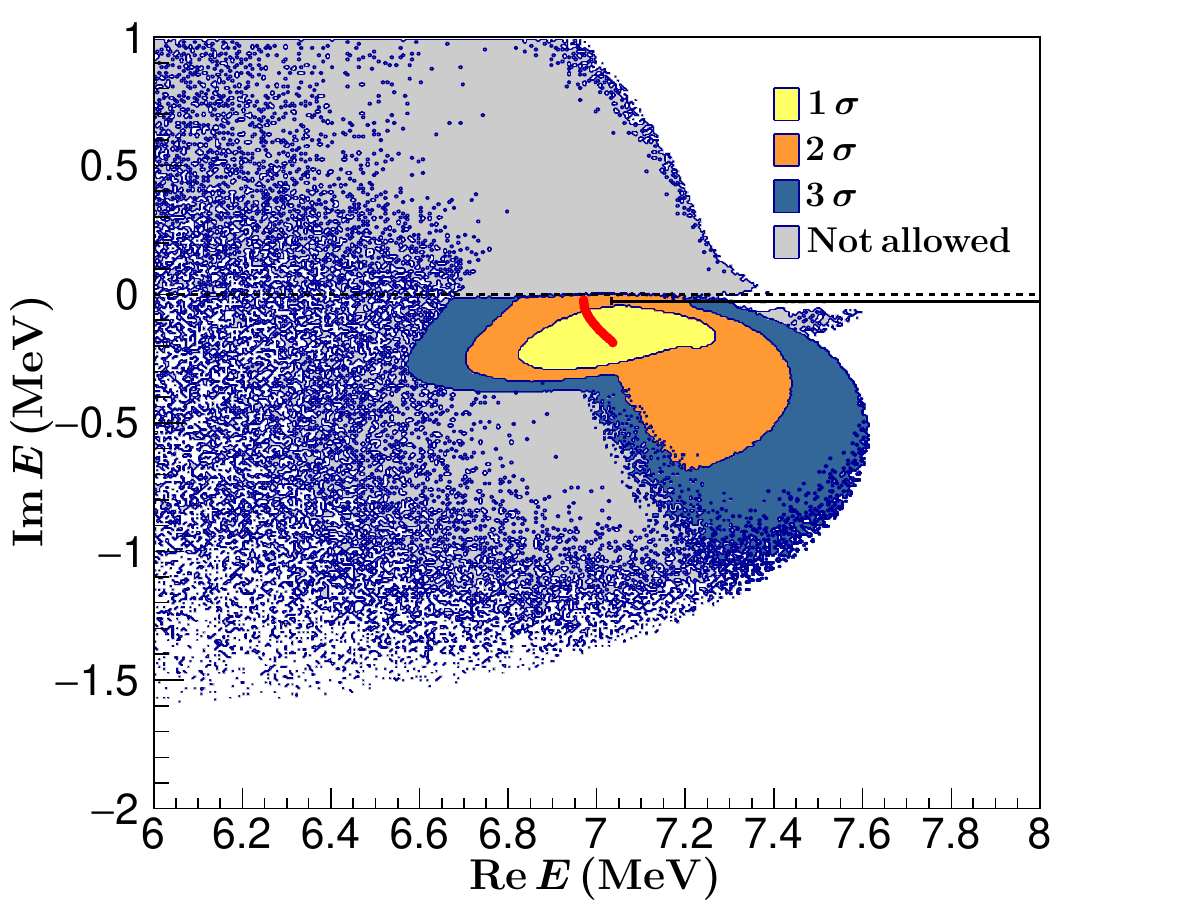}}%
\subfigure[~sheet \uppercase\expandafter{\romannumeral2}: $0.26-1.71i$ MeV]{
\includegraphics[height=4.6cm]{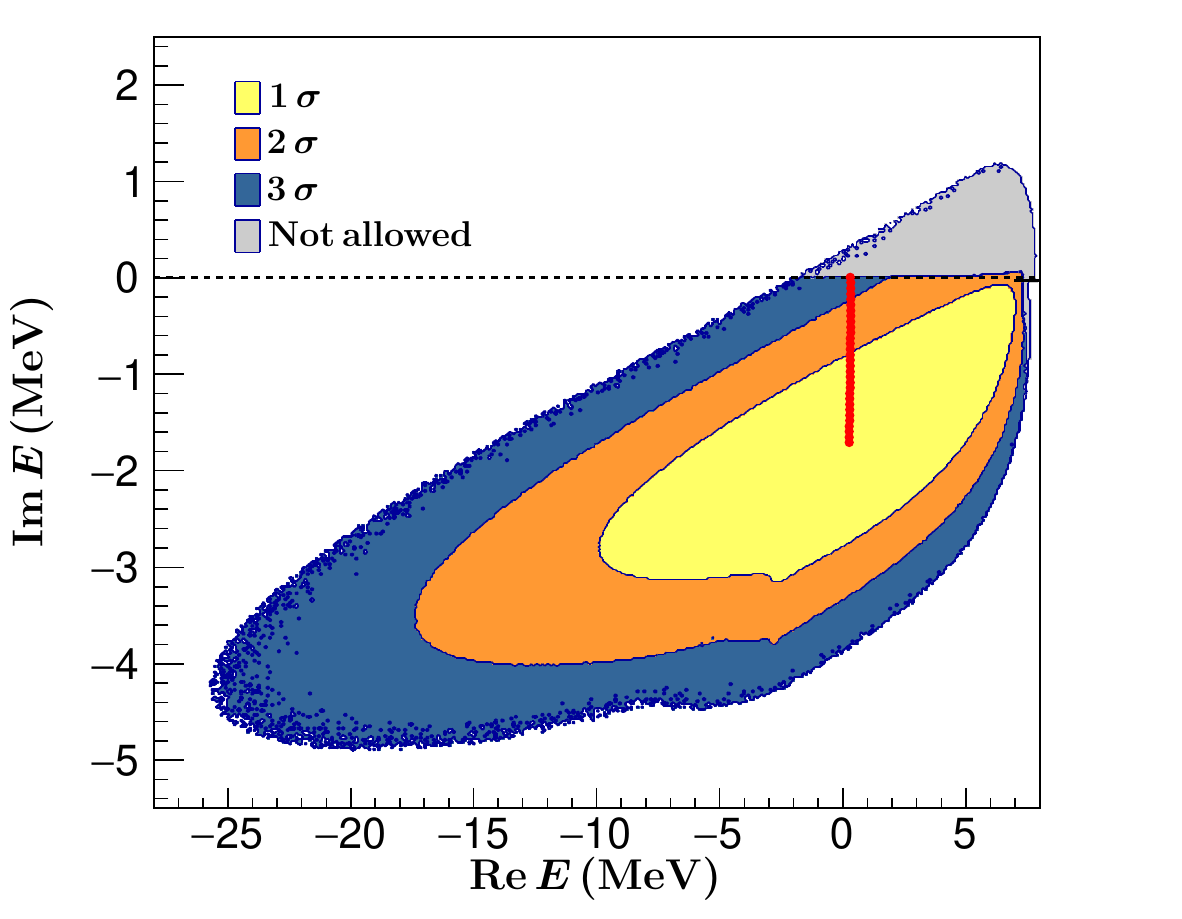}}%
\subfigure[~$k$-plane]{
\includegraphics[height=4.6cm]{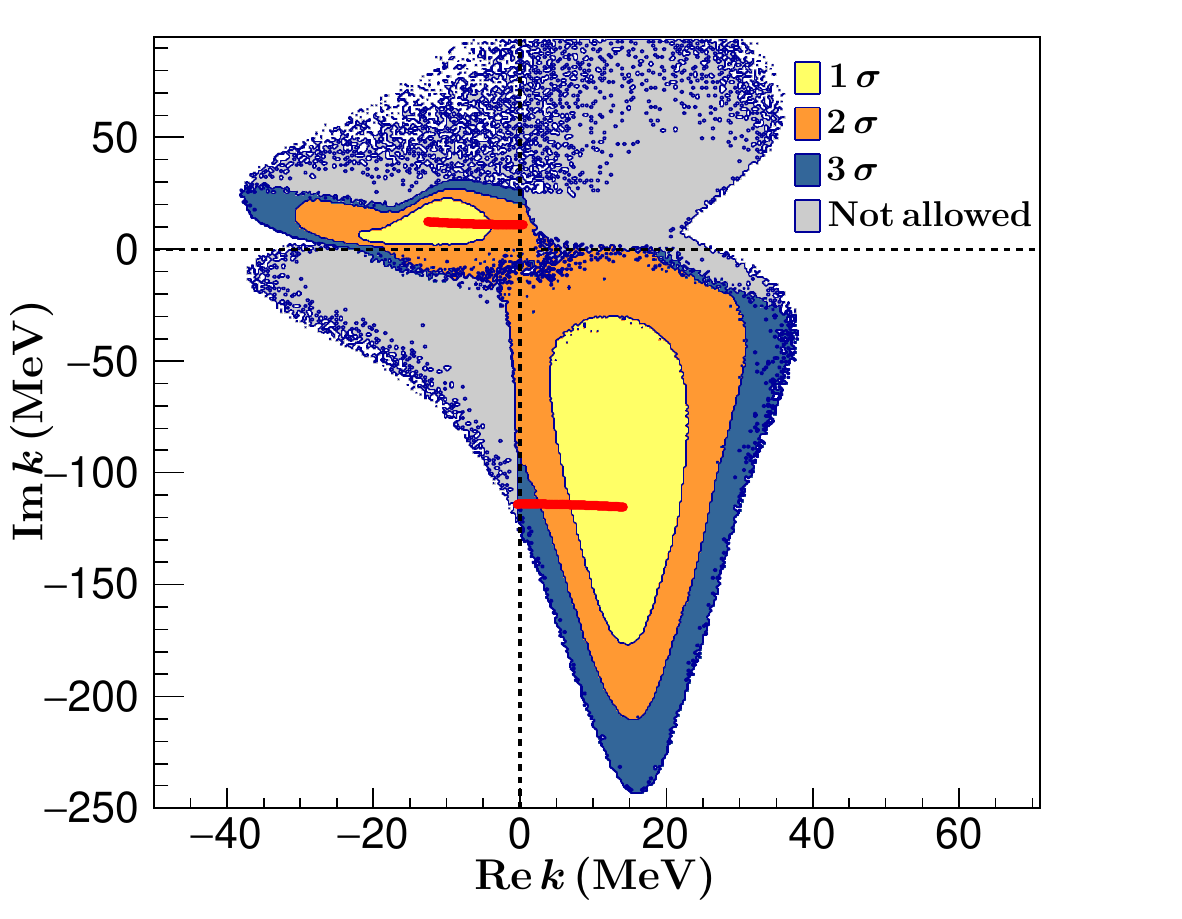}}
\caption{\label{fig:ncoutour} Confidence regions of the pole position on (a) sheet \uppercase\expandafter{\romannumeral1} and (b) sheet \uppercase\expandafter{\romannumeral2}. Only the statistical uncertainty is shown. The gray region labeled ``Not allowed'' indicates the region where $g$ or $\Gamma_{0}$ is negative. (c) Confidence region of the pole positions on the $k$-plane.}
\end{figure}

The confidence regions of the pole locations are shown in Fig.~\ref{fig:ncoutour}, and the covarant matrix of the pole locations are shown in Table~\ref{tab:covPole} and \ref{tab:covK}. The systematic uncertainties are summarized in Table~\ref{Tab:errPole} and \ref{Tab:errPoleK}.

\begin{table}[htbp]
\renewcommand\arraystretch{1.2}
\caption{\label{tab:covPole}The covariance matrix of the locations of the two poles on the two Riemann sheets.}
\vspace{8pt}
\centering 
\begin{tabular}{ccccc}
\hline\hline
&Re[$\Eone$] (${\rm MeV}$)&Im[$\Eone$] (${\rm MeV}$)&Re[$\Etwo$] (${\rm MeV}$)&Im[$\Etwo$] (${\rm MeV}$)\\\hline
Re[$\Eone$]&0.02&-&0.53&-0.01\\
Im[$\Eone$]&&0.01&0.22&0.03\\
Re[$\Etwo$]&&&32.96&3.79\\
Im[$\Etwo$]&&&&0.82\\
\hline\hline
\end{tabular}
\end{table}

\begin{table}[htbp]
\renewcommand\arraystretch{1.2}
\caption{\label{tab:covK}The covariance matrix of the locations of the two poles on the $k$-plane.}
\vspace{8pt}
\centering \begin{tabular}{ccccc}
\hline\hline
&Re[$k^+$] (${\rm MeV}$)&Im[$k^+$] (${\rm MeV}$)&Re[$k^-$] (${\rm MeV}$)&Im[$k^-$] (${\rm MeV}$)\\\hline
Re[$k^+$]&33.4&16.7&-31.4&-53.6\\
Im[$k^+$]&&46.3&-14.3&-190.3\\
Re[$k^-$]&&&33.8&8.3\\
Im[$k^-$]&&&&1986.3\\
\hline\hline
\end{tabular}
\end{table}

\begin{table}[t]
\renewcommand\arraystretch{1.2}
\tabcolsep=0.2cm
\caption{\label{Tab:errPole}The upper and lower uncertainties on the pole locations on the Riemann sheets.}
\vspace{8pt}
\centering 
\begin{tabular}{lllll}
\hline\hline
Source&\multicolumn{1}{c}{Re[$\Eone$]~(MeV)}&\multicolumn{1}{c}{Im[$\Eone$]~(MeV)}&\multicolumn{1}{c}{Re[$\Etwo$]~(MeV)}&\multicolumn{1}{c}{Im[$\Etwo$]~(MeV)}\\\hline
$\alpha$&\multicolumn{1}{c}{$\pm0.01$}&$+0.12~-0.18$&$+4.74~-34.20$&$+0.54~-0.76$\\
$\Gamma_{D^{*0}}$&\multicolumn{1}{c}{$\pm0.01$}&\multicolumn{1}{c}{$\pm0.01$}&$+0.23~-0.20$&\multicolumn{1}{c}{$\pm0.03$}\\
Efficiency&\multicolumn{1}{l}{$+0.01$}&\multicolumn{1}{c}{$\pm0.02$}&$+1.55~-2.41$&$+0.13~-0.18$\\
Resolution&\multicolumn{1}{c}{$-$}&\multicolumn{1}{c}{$-$}&\multicolumn{1}{c}{$\pm0.08$}&\multicolumn{1}{c}{$\pm0.01$}\\
Background&\multicolumn{1}{c}{$\pm0.02$}&\multicolumn{1}{l}{$+0.01$}&$+0.15~-3.02$&$\phantom{+0.12~}-0.26$\\
$M(D^0)$ &\multicolumn{1}{c}{$\pm0.06$}&\multicolumn{1}{c}{$\pm0.03$}&$+0.34~-0.16$&$+0.12~-0.11$\\
$E_{\rm cms}$&$\phantom{+0.12~}-0.03$&$\phantom{+0.12~}-0.06$&$\phantom{+0.12~}-16.79$&$\phantom{+0.12~}-1.77$\\
Simulation&\multicolumn{1}{c}{$\pm0.01$}&\multicolumn{1}{c}{$\pm0.01$}&\multicolumn{1}{c}{$\pm1.15$}&\multicolumn{1}{c}{$\pm0.18$}\\\hline
Sum&$+0.07~-0.08$&$+0.14~-0.19$&$+5.14~-38.32$&$+0.60~-1.96$\\
\hline\hline
\end{tabular}
\end{table}

\begin{table}[t]
\renewcommand\arraystretch{1.2}
\tabcolsep=0.2cm
\caption{\label{Tab:errPoleK}The upper and lower uncertainties on the pole locations on the $k$-plane.}
\vspace{8pt}
\centering \begin{tabular}{lllll}
\hline\hline
Source&\multicolumn{1}{c}{Re[$k^+$]~(MeV)}&\multicolumn{1}{c}{Im[$k^+$]~(MeV)}&\multicolumn{1}{c}{Re[$k^-$]~(MeV)}&\multicolumn{1}{c}{Im[$k^-$]~(MeV)}\\\hline
$\alpha$&$+5.9~-6.0$&$+5.6~-5.9$&$+4.9~-1.6$&$+49.8~-162.5$\\
$\Gamma_{D^{*0}}$&$\phantom{+5.9~}-0.2$&$+0.7~-0.5$&$\phantom{+5.9~}-0.1$&$+2.0~-1.7$\\
Efficiency&\multicolumn{1}{c}{$\pm1.0$}&$+0.6~-0.7$&$+0.6~-0.7$&$+13.7~-18.6$\\
Resolution&\multicolumn{1}{c}{$-$}&\multicolumn{1}{c}{$-$}&\multicolumn{1}{c}{$-$}&\multicolumn{1}{c}{$\pm0.7$}\\
Background&\multicolumn{1}{c}{$\pm1.0$}&$+0.5~-0.6$&$+0.5~-1.0$&$+1.1~-24.3$\\
$M(D^0)$ &$+0.6~-0.7$&$+1.7~-1.9$&$+0.8~-0.7$&$+2.6~-1.0$\\
$E_{\rm cms}$&\multicolumn{1}{l}{$+2.7$}&\multicolumn{1}{l}{$+0.2$}&\multicolumn{1}{l}{$+1.5$}&$\phantom{+15.9~}-98.7$\\
Simulation&\multicolumn{1}{c}{$\pm0.3$}&\multicolumn{1}{c}{$\pm0.8$}&\multicolumn{1}{c}{$\pm0.4$}&\multicolumn{1}{c}{$\pm9.9$}\\\hline
Sum&$+6.6~-6.2$&$+6.0~-6.4$&$+5.3~-2.1$&$+52.7~-192.8$\\
\hline\hline
\end{tabular}
\end{table}